\documentclass[11pt]{article}

\usepackage [latin1]{inputenc}
\usepackage{ifpdf}
\usepackage{amsmath}
\usepackage{graphicx}
\usepackage{indentfirst}
\usepackage{amssymb}
\usepackage{cite}
\usepackage{color}
\usepackage{subfigure}
\usepackage[colorlinks,linkcolor=blue,citecolor=blue,urlcolor=blue,hyperindex]{hyperref}
\usepackage{indentfirst}
\usepackage{amsmath}
\usepackage{latexsym}
\usepackage{amsmath}
\usepackage{color}
\usepackage{lineno}
\usepackage{graphicx}
\usepackage{float}
\usepackage{subfigure}
\usepackage{geometry}
\usepackage{graphics}
\usepackage{subfigure}
\usepackage{graphicx}
\usepackage{multirow}
\usepackage{booktabs}
\usepackage{hyperref}
\usepackage{cite}
\usepackage{wasysym}

\begin{document}	
\title{The shadow and observational images of the non-singular rotating black holes in loop quantum gravity}
	
\date{}
\maketitle
	
\begin{center}
		\author{Guo-Ping Li}$^{a,}$$\footnote{\texttt{Corresponding author:gpliphys@yeah.net}}$,
		\author{He-Bin Zheng}$^{a,}$$\footnote{\texttt{zhenghb3060@163.com}}$,
        \author{Ke-Jian He}$^{b,}$$\footnote{\texttt{kjhe94@163.com}}$ and
		\author{Qing-Quan Jiang}$^{a,}$$\footnote{\texttt{Corresponding author:qqjiangphys@yeah.net}}$,
		\vskip 0.25in

$^{a}$\it{Physics and Space College, China West Normal University, Nanchong 637000, People's Republic of China}\\
$^{b}$\it{School of Civil Engineering, Chongqing Jiaotong University, Chongqing 400074, People's Republic of China}\\
		
\end{center}
\vskip 0.6in
\begin{abstract}
By considering the celestial light source and the thin disk source, we employ the backward ray-tracing method to carefully study the shadow, inner shadow and observational images of the non-singular rotating black holes in loop quantum gravity.
The results show that the increase of quantum parameter $\lambda$ causes the shadow to shrink, while increases the deviation from circularity. And, the shadow's angular diameter of M87* impose stronger constraints on the observed properties of the no-singulgar rotating black holes by comparing with SgrA*.
For a celestial light source, the parameter $\lambda$ indeed influences the distortion of light around black hole shadow, but this effect is relatively small and only becomes noticeable when extremely close to the shadow.
When a thin accretion disk around black hole, it turns out that for an observer at any position, the parameter $\lambda$ has little effect on the shape of the inner shadow. However, it decreases the size of the inner shadow, reduces the observed light intensity, and narrows the redshifted shadow images, regardless of whether the accretion disk is prograde or retrograde.
Meanwhile, it is true that the thin disk images of black hole cannot effectively reflect the internal structure of black hole.
Finally, we can conclude that a key observational feature of these non-singular rotating black holes is that the larger the black hole's spin parameter, the smaller the upper limit of $\lambda$'s effect.
And, the parameter $\lambda$ decreases the gravitational field's strength, thereby weakens the observed images.
This could provide a possible way to constraining black hole parameters, identifying quantum gravity effects, and distinguishing loop quantum gravity black holes, even if it cannot be used to distinguish the non-singular properties of black hole.
\end{abstract}
	
\thispagestyle{empty}
\newpage
\setcounter{page}{1}
	
\section{Introduction}\label{sec1}

Since 2019, the Event Horizon Telescope (EHT) international collaboration team has obtained images of the black hole at the center of the radio galaxy M87* and the Milky Way's SgrA* \cite{EventHorizonTelescope:2019dse,EventHorizonTelescope:2022wkp}. Both images share similarities, featuring a relatively dark central region and a brighter outer ring structure. These two regions are referred to as the black hole shadow and the photon ring, respectively. In fact, light near a black hole is absorbed by the black hole and cannot reach the observer, resulting in a dark region against the background sky when observing the black hole \cite{Cunha:2018acu}. This dark region is the black hole shadow. Strictly speaking, due to the deflection effects of the gravitational field on light, the dark region within the critical curve is defined as the black hole shadow, while light approaching the critical curve from infinity will asymptotically approach the boundary photon orbit \cite{Perlick:2021aok}. The advent of black hole images marks a significant milestone in the history of black hole research and holds substantial scientific significance. These images not only provide valuable information about the accretion processes, radiation mechanisms, and jet mechanisms near black holes, but also extract the spacetime characteristics of black holes. For a long time, the black hole shadow and its observable effects have been one of the focal points of research. For example, the shadow of a Schwarzschild black hole appears as a black disk from any angle, while an observer in the equatorial plane will notice that the shadow of a Kerr black hole changes shape with the rotation parameter. As the rotation parameter increases, the black disk gradually evolves into a ``D" shape \cite{Chandrasekhar}.
Later on, detailed studies have been conducted on the shadows of rotating EMDA black hole \cite{Wei:2013kza}, non-commutative black holes \cite{Wei:2015dua}, phantom black holes \cite{Huang:2016qnl}, Konoplya-Zhidenko black holes \cite{Cunha:2017eoe}, and black hole with massive vector fields \cite{Wang:2017hjl}, revealing the double shadows, cuspy shadow structures and other interesting phenomenon \cite{He:2024qka,Guo:2020zmf,Atamurotov:2013sca,Perlick:2015vta,Konoplya:2019sns,Shaikh:2018lcc,Abdujabbarov:2016hnw,Amarilla:2011fx,Nedkova:2013msa,Papnoi:2014aaa,Meng:2023wgi}. Based on black hole shadow, one has explored various aspects of black hole, i.e., the quasinormal modes and parameter constraints of black holes, among other aspects, which offer deep insights into our understanding and comprehension of gravity \cite{Younsi:2016azx, Konoplya:2019xmn, Konoplya:2024lch, Konoplya:2021slg,Shaikh:2021yux, Chowdhuri:2020ipb, Soares:2023uup,Aliyan:2024xwl,Soares:2023err,Soares:2024rhp,Olmo:2023lil,
Tsukamoto:2014tja,Tsukamoto:2017fxq, Ghosh:2022gka,Nozari:2024jiz,Nozari:2023flq,Pal:2023wqg}.
To capture a more detailed view of black hole shadows and the light distortion behavior around them, based on a four-color celestial sphere light source model, one has numerically studied Einstein rings caused by gravitational lensing effects, as well as images of black hole shadows and the space-dragging effect caused by rotation \cite{Cunha:2015yba}. This method was later extended to black hole models in various gravitational theories, used for numerically simulating black hole shadows when the photon motion system is integrable \cite{Chen:2022scf}. In fact, black holes in the universe are always surrounded by various accreting materials \cite{Luminet:1979nyg}. In view of this, when a simple static optically thin spherical accretion surrounds a Schwarzschild black hole, Narayan et al. discovered in 2019 that the black hole shadow and photon sphere reflect the geometry of spacetime, and their sizes are almost unaffected by the accreting material \cite{Narayan:2019imo}.
In fact, there is another typical disk-shaped accreting matter exists in the universe, namely the accretion disk. And, the astronomical observations indicate that the accretion disk indeed exists around the supermassive black hole SgrA* at the center of the Milky Way. In 2019, Wald et.al discovered that when a Schwarzschild black hole is surrounded by the accretion disk (the geometrically and optically thin disk), the observer located at the North pole would find the black hole's shadow as a dark disk encircled by a bright ring. This bright ring is composed of direct emissions, a lensing ring, and a photon ring \cite{Gralla:2019xty}. Later on, this idea was further extended to other static spherically symmetric black holes containing matter fields and modified grvities, or to boson stars or wormhole spacetimes, and so so on 
\cite{Zeng:2020vsj,Li:2021riw,Zeng:2021mok,Zeng:2021dlj,Zeng:2022pvb,Zeng:2022fdm,Rosa:2022toh,Asukula:2023akj,Rosa:2022tfv,Rosa:2024eva,Li:2021ypw,Peng:2020wun,Peng:2021osd,Zhang:2023okw,Meng:2024puu,Meng:2023htc,Gan:2021xdl}. It was proposed that the structure of the black hole shadow and the associated rings could be used to distinguish between different black hole models in various gravitational theories.
Considering the strong magnetic field around black holes, in 2022, this work \cite{Gralla:2019xty} was extended for the first time to the case of rotating Kerr black holes, analyzing the effects of rotation, magnetic fields, and the observer's angle on the multi-level images of black holes \cite{Hou:2022eev}. In \cite{Hou:2022eev}, it was assumed that the accretion disk around the Kerr black hole is composed of plasma fluid and is located in the equatorial plane of black hole. The fluid moves along Keplerian orbits outside the innermost stable circular orbit (ISCO), while inside the ISCO, it moves along simplified precessing geodesic trajectories with the energy and angular momentum of the ISCO. Of course, in recent years, some progresses have been made in the study of black hole shadows and images, particularly in exploratory research on hotspot images \cite{Huang:2024wpj,Chen:2023knf}, polarized images \cite{Lee:2022rtg,Qin:2022kaf,Chen:2024cxi}, jet images \cite{Zhang:2024lsf}, boson star images \cite{Rosa:2023qcv}, accretion disk models \cite{Vincent:2022fwj}, and so on. At present, by considering the thin disk models \cite{Hou:2022eev}, it should be noted that the rotating black hole images in other gravity models and matter fields remains unknown. Therefore, continuing to study black hole images under this model would be a highly significant endeavor, as it may provide an effective tool for distinguishing different gravitational models and exploring quantum effects.

On the other hand, black hole singularitity is the point in spacetime where the curvature becomes infinite, at which the mass of matter is compressed infinitely, with the density approaching to infinity and the volume approaching zero. And, Penrose have proven that all matter within a black hole should collapse into the singularity \cite{Penrose:1964wq}. The formation of singularities is almost inevitable within the framework of classical general relativity, making it impossible to predict the evolution of physical phenomena in spacetime effectively. Currently, it is widely believed that the existence of spacetime singularities reflects the incompleteness of general relativity, and that resolving the singularity problem likely requires a theory of quantum gravity(QG). To date, a complete theory describing QG has yet to be found. This has prompted physicists to focus on constructing nonsingular black holes within classical general relativity. Typically, one can seek nonsingular solutions to the field equations to address the singularity problem, at the cost of violating the strong energy condition. These solutions lack intrinsic singularities at the coordinate origin, and event horizons can still exist. Such black holes are referred to as non-singular black holes or non-singular black holes. In 1968, Bardeen proposed the first nonsingular black hole, namely Bardeen black hole, which satisfies the weak energy condition \cite{Bardeen}. In 2000, it was shown that the physical motivation for the absence of spacetime singularities in this solution is that the Bardeen black hole is a spherically symmetric, static solution derived from coupling a nonlinear electrodynamics model to Einstein's gravitational field equations \cite{Ayon-Beato:2000mjt}.
Subsequently, many similar spherically symmetric non-singular black holes have been constructed, such as the Hayward black hole, the Ay\'{o}n-Beato-Garc\'{i}a black hole \cite{Ayon-Beato:1998hmi}, the Berej-Matyjasek-Trynieki-Wornowicz black hole \cite{Berej:2006cc}, and the Simpson-Visser black hole \cite{Simpson:2018tsi}, among others \cite{Hayward:2005gi,Ling:2021olm,Bronnikov:2005gm,Bambi:2013ufa,Fan:2016hvf,Balart:2014cga,Azreg-Ainou:2014pra,Johannsen:2013szh,Toshmatov:2014nya}. Since non-singular black holes are considered manifestations of QG effects, various properties of non-singular black holes have been extensively studied, such as quasinormal modes \cite{Fernando:2012yw}, greybody factors \cite{Jusufi:2020agr}, superradiance \cite{Yang:2022uze}, Joule-Thomson expansion \cite{Rajani:2020mdw}, P-V criticality \cite{Singh:2020xju}, dynamical stability \cite{Moreno:2002gg}, gravitational lensing \cite{Ramadhan:2023ogm}, shadow profiles \cite{Zare:2024dtf}, and spherical accretion shadow appearances \cite{He:2021htq}, among others. By exploring these properties, it is possible for us to capture effects related to QG. Currently, the study of non-singular black holes has become one of the focal points and hot topics in contemporary research.\\

The loop quantum gravity(LQG), as one of the main candidates for QG theory, is a non-perturbative approach to QG that introduces new dynamical variables in a connection dynamics framework, which has drawn a great deal of attention \cite{Rovelli:1997yv,Meissner:2004ju,Han:2005km,Yang:2009fp,Zhang:2011vi, Papanikolaou:2023crz, Ma:2010fy,Chen:2024sbc,Jiang:2023img,Makinen:2023shj,Long:2021lmd}. This theory not only has a well-defined Hilbert space for LQG dynamics but also naturally predicts a discrete structure of spacetime geometry at the Planck scale. More importantly, the non-rotating LQG black hole solutions obtained in this gravitational theory exhibits a non-singular geometry that we expect due to the existence of transition surface. By using the Newman-Janis-Algorithm(NJA) that is a solution-generating method \cite{Newman:1965tw}, one has constructed the rotating black hole solutions from the Schwarzschild black hole \cite{Brahma:2020eos}. This solution does not depend on the specific details of the seed metric used, thereby effectively capturing some universal properties of rotating LQG black holes. At present, the dynamical behavior of a scalar field near this black hole has been investigated \cite{Xia:2023zlf}. However, it remains unclear whether the non-singular effects of this black hole (or the influence of the transition surface) and quantum effects have observable astronomical consequences.
Therefore, based on the current progress in black hole images, it is a highly important research for studying the shadow, inner shadow, celestial source images, and thin disk images of non-rotating LQG black holes by using the thin disk accretion model. In this context, this paper will focus on the astronomical observable effects of non-rotating LQG black holes, providing potential references for distinguishing non-singular black holes and analyzing the quantum effects of black holes.

The structure of paper is as follows: In Section \ref{sec2}, we briefly review the non-singular rotating black holes in LQG. In Section \ref{sec3}, we study the shadow of the black hole, and analyzed the deviation from the circularity and the size of shadow; Also, the shadow angular diameter is obtained, and compared with that of M87* and SgrA*. In section \ref{sec4}, the images of black hole with the celestial light source are presented. Section \ref{sec5} is devoted to investigate the images illuminated by the thin accretion disk.
Finally, in section \ref{sec7}, we give a brief conclusion and discussion.

\section{Review of the rotating LQG black holes}\label{sec2}
\begin{figure}[!h]
    \centering
    \includegraphics[width=0.4\linewidth]{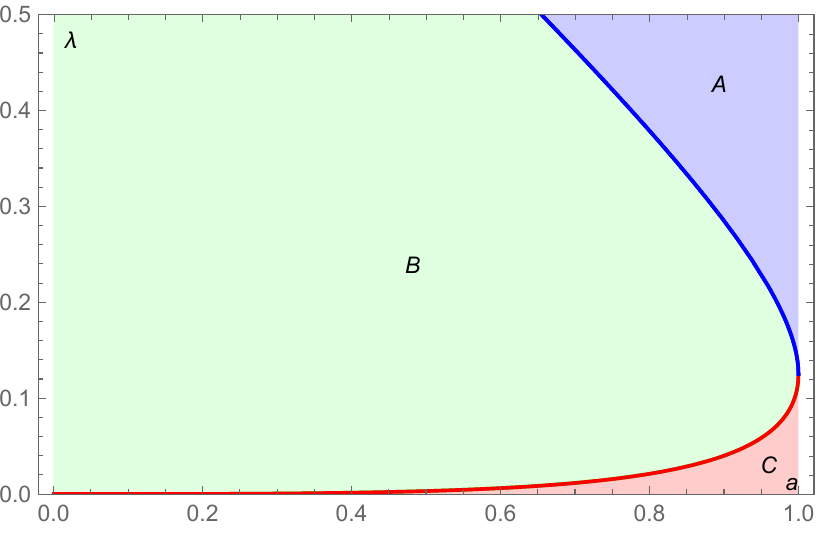}
    \caption{The spacetime structure of the non-singular rotating black holes in LQG.}
    \label{fig1}
\end{figure}

In this section, we briefly review the non-singular rotating black holes in LQG.
Because of the complexity of Einstein's field equations in the rotating case, the Newman-Janis algorithm(NJA) method is widely accepted during the construction of the rotating black holes.
The solution of the rotating LQG black holes we focused on in this paper is obtained by this method.
In general, the seed metric in a general static and spherically symmetric space time in 4-dimensional case always reads
\begin{equation}\label{eq1}
ds^2=- g(r) d t^2 + \frac{dr^2}{f(r)} + h(r) (d\theta^2 +\sin \theta^2 d\phi^2 ).
\end{equation}
Based on the NJA, by introducing the advanced null coordinates $(u,t,\theta,\phi)$ where $d u= d t - \frac{dr}{\sqrt{f(r) g(r)}}$, one can use a null tetrad ($Z_{a}^{\mu} = (l^{\mu}, n^{\nu}, m^{\mu}, \bar{m}^{\mu})$) to express the inverse of the seed metric
\begin{equation}\label{eq2}
g^{\mu \nu}= -l^{\mu} n^{\nu} - l^{\nu} n^{\mu} + m^{\mu} \bar{m}^{\nu} + m^{\nu} \bar{m}^{\mu} ).
\end{equation}
In Eq.(\ref{eq2}), $\bar{m}^{\mu}$ is the complex conjugate of ${m}^{\mu}$. The null tetrad in the advanced null coordinates can be expressed as following
\begin{align}\label{eq3}
&l^{\mu} = \delta_r^{\mu}, \quad n^{\mu} = \sqrt{\frac{f(r)}{g(r)}}\delta_{\mu}^{\mu}-\frac{f(r)}{2} \delta_r^{\mu}, \quad m^{\mu} =  \frac{1}{\sqrt{2 h(r)}} \left( \delta_{\theta}^{\mu} +\frac{i}{\sin \theta} \delta_{\phi}^{\mu}  \right),
\end{align}
and they should satisfy
\begin{align}\label{eq3a1}
&l_{\mu} l^{\mu} = m_{\mu} m^{\mu} = n_{\nu} n^{\nu} = l_{\mu} m^{\mu} =n_{\mu} m^{\mu} =0,  \quad l_{\mu} n^{\mu}  = - m_{\mu} \bar{m}^{\mu} =1.
\end{align}
To introduce the spin in NJA, one should perform a complex shift on the advanced null coordinates as
\begin{align}\label{eq4}
r \rightarrow r + i a \cos \theta, \quad u \rightarrow u - i a \cos \theta,
\end{align}
with $a$ is the rotation parameter. Under this transformation, the $\delta_{\nu}^{\mu}$ transform as a vector, which is
\begin{align}\label{eq5}
\delta_r^{\mu} \rightarrow \delta_r^{\mu} , \quad  \delta_u^{\mu} \rightarrow \delta_u^{\mu},   \quad  \delta_{\theta}^{\mu} \rightarrow  \delta_{\theta}^{\mu}+ i a \sin \theta ( \delta_{u}^{\mu} -  \delta_{r}^{\mu}),  \quad  \delta_{\phi}^{\mu} \rightarrow   \delta_{\phi}^{\mu}.
\end{align}
In the context of above complex transformation, the metric functions would generically be functions of $(r, \theta, a)$. But, this transformation does not work well as discussed in \cite{Lan:2023cvz}. So, one denotes the metric
functions after the complex shift by $\{g(r,\theta,a), f(r,\theta,a), h(r,\theta,a) \}$$\rightarrow$ $\{A(r,\theta,a), B(r,\theta,a), \Psi(r,\theta,a)\}$. Here, $A(r,\theta,a), B(r,\theta,a), \Psi(r,\theta,a) $ are all the real function, and which should recover their static counterparts in the limit $a \rightarrow 0$. In this consideration, the corresponding metric with inclusion of the rotation can be expressed as
\begin{align}\label{eq6}
ds^2 &= - Adu^2 -2 \sqrt{\frac{A}{B}} du dr -2 a \sin^2 \theta \left( \sqrt{\frac{A}{B}} -A  \right) du d\phi +2 a \sin^2 \theta \sqrt{\frac{A}{B}} dr d\phi + \Psi d \theta^2 \nonumber \\
&+ \sin^2 \theta \left[\Psi + a^2 \sin^2 \theta \left(
   \sqrt{\frac{A}{B}} -A  \right)      \right] d\phi^2.
\end{align}
With the help of the coordinate transformation and some constraints in \cite{Lan:2023cvz}, we can return to the Boyer-Lindquist coordinates, and can only considered one off-diagonal term $g_{t \phi}$ in above metric. Finally, when one reconsidered the seed metric in LQG that described in \cite{Yang:2023cmv,Brahma:2020eos}, the rotating black holes with the Kerr-like form can be expressed as
\begin{equation}\label{eq7}
ds^2=-\left(1-\frac{F(r) \mathcal{H}(r)}{\Sigma}\right)dt^2+ \frac{\Sigma}{\Delta}dr^2  + \Sigma d \theta ^2 +\frac{\mathcal{A} \sin^2\theta}{\Sigma}  d \phi^2  -\frac{2 F(r) \mathcal{H}(r) a }{\Sigma} \sin^2{\theta} dt d\phi,
\end{equation}
with
\begin{align}\label{eq8}
&\Sigma=\mathcal{H}(r) + a^2 \cos^2{\theta}, \nonumber\\
&F(r)=1-\mathcal{F}(r), \nonumber \\
&\Delta=\mathcal{F}(r) \mathcal{H}(r)+ a^2, \nonumber \\
&\mathcal{A}=\left(\mathcal{H}(r)+a^2\right)^2 - a^2 \Delta \sin^2{\theta},
\end{align}
and,
\begin{align}\label{eq9}
&\mathcal{H}(r)=r^2 + \lambda^{2/3} M^{2/3}, \nonumber\\
&\mathcal{F}(r)=\left(1-\frac{2 M}{\sqrt{r^2 + 4\lambda^{2/3}M^{2/3}}}\right) \frac{r^2 + 4\lambda^{2/3}M^{2/3}}{r^2 + \lambda^{2/3}M^{2/3}}.
\end{align}
In fact, $\Psi(r,\theta,a)$ should be determined by some especial physical interpretations. However, here when the source is interpreted as an imperfect fluid rotating about the $z$ axis, $\Psi(r,\theta,a)$ is obtained, which is $\Psi(r,\theta,a) = \Sigma=\mathcal{H}(r) + a^2 \cos^2{\theta} $. The quantum parameter $\lambda$ originates from holonomy modifications, which represents the effects of QG.
For the non-singular rotating black hole (\ref{eq7}), the horizons reads
\begin{align}\label{eq10}
&\Delta=\mathcal{F}(r) \mathcal{H}(r)+ a^2 = r^2 + 4\lambda^{2/3}M^{2/3} + 2M \sqrt{r^2 + 4\lambda^{2/3}M^{2/3}} +a^2=0.
\end{align}
We have
\begin{align}\label{eq11}
&r_{\pm}=\sqrt{\left(M \pm \sqrt{M^2 - a^2}\right)^2 - 4 \lambda^{2/3} M^{2/3}},
\end{align}
where, the symbol $\pm$ denotes the outer and inner horizons of black hole, respectively.
Also, for the LQG black hole, there always exists a transition surface which means the areal radius reaches to a minimum value. For different choices of $(a, \lambda)$, the horizon of (\ref{eq7}) can be very different, which can be see from Fig.\ref{fig1}.

In parameter space $(a, \lambda)$, it shows that there are three regions, namely $A, B, C$. Region A, i.e., the blue region in Fig.\ref{fig1}, represents a rotating wormhole with no horizon existed, where the transition surface is outside of the outer horizon. Region B, i.e., the light green region, denotes the rotating black hole with only an outer horizon, where the transition surface is located between the inner and outer horizons. Region C, i.e., the red region, is the non-singular rotating black hole with the transition surface is located inside the inner horizon.
By using the shadow size of M87* measured by EHT, one has argued that the possibility of the rotating LQG compact object being a wormhole without horizon has been almost ruled out \cite{Brahma:2020eos}. In view of this, we will employ the some acceptable parameters in $B$ and $C$ to carefully study the shadow and observable appearance of the non-singular rotating black hole in the context of LQG. In particular, we mainly focus on the effects of quantum parameter $\lambda$ and rotating parameter $a$ on the shadow and appearance of the non-singular black hole.

\section{Shadows of the rotating LQG black holes}\label{sec3}
\begin{figure}[!h]
    \centering
\includegraphics[width=0.4\linewidth]{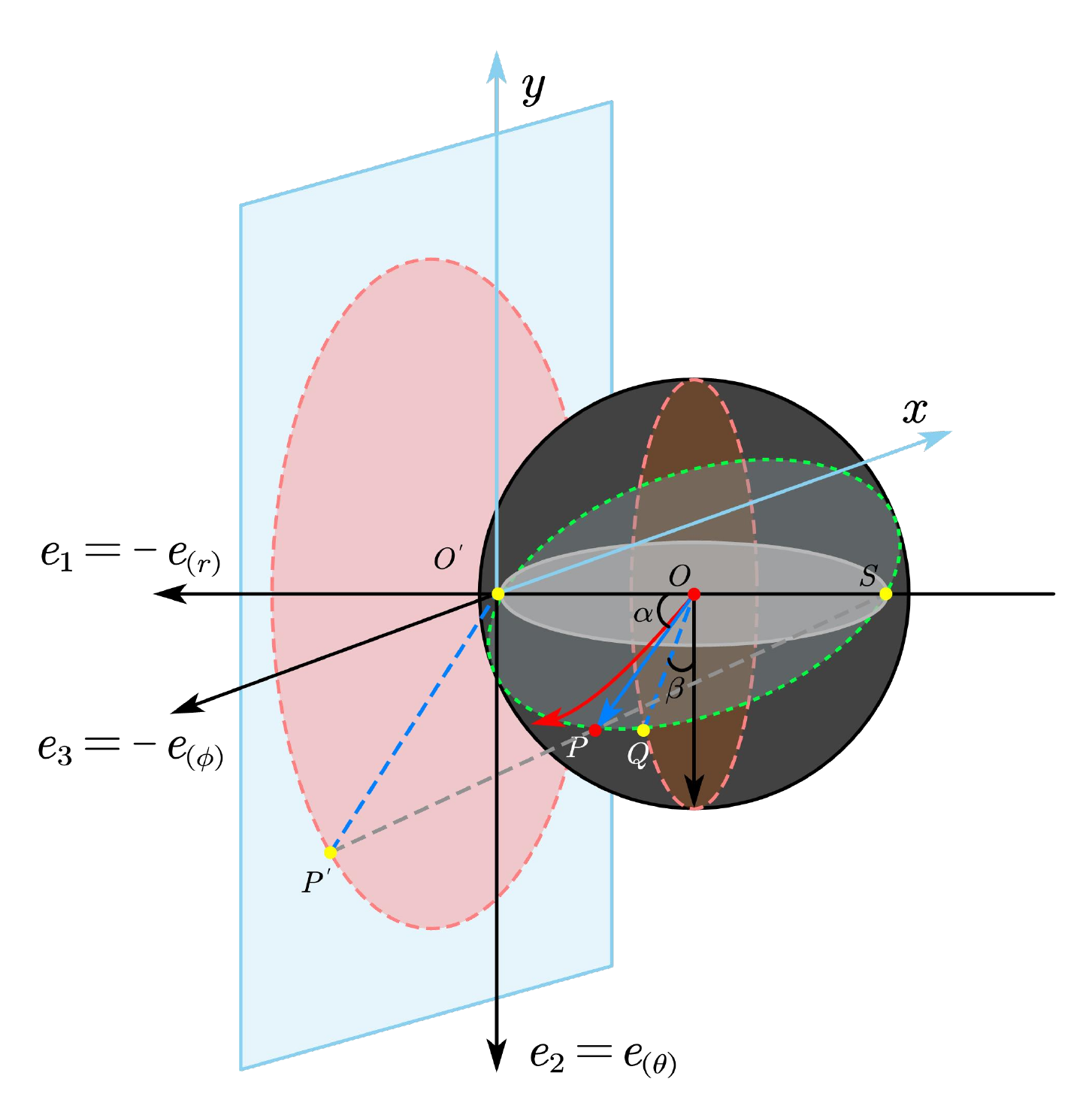}
    \caption{The ZAMO tetrad and celestial coordinates $\alpha$ and $\beta$ based on the method of stereographic projection \cite{Hu:2020usx}.}
    \label{fig2}
\end{figure}

Starting from the geodesics of the photons, we will study the shadow of non-singular rotating black holes in LQG.
To study the geodesics of particles with mass $m$, here we introduce
\begin{align}\label{eq12}
H = \frac{1}{2} g_{\mu \nu} p_{\mu}p_{\nu} = -\frac{m^2}{2},
\end{align}
where $H$ is the canonical Hamiltonian, $p_{\mu}$ is the four momentum.
By considering the symmetries of the spacetime and the associated Killing vectors,
there are two conserved quantities along a geodesic, namely, $E$ and $L$, which represent the energy and angular momentum in the direction of the axis of symmetry. They are
\begin{align}\label{eq13}
E:= -p_t =  - g_{\phi t} \dot{\phi} - g_{tt} \dot{t}, \quad L:= p_{\phi} = g_{\phi \phi} \dot{\phi} +  g_{\phi t} \dot{t}.
\end{align}
The symbol $\dot{t}$ represents the derivative of $\tau$, which is the affine parameter.
For null geodesics, i.e., $m=0$, the equations of motion for photons propagating in the spacetime (\ref{eq7}) can be obtained by solving the Hamilton-Jacobi equation. Further combined with Eq.(\ref{eq13}), the equations of motion for photons can be expressed as the following four first-order differential equations
\begin{align}\label{eq16}
& \Sigma \dot{t}={E} \Sigma  + \frac{\left(a^2-\Delta +\mathcal{H} \right) \left[{E} \left(a^2+\mathcal{H} \right)-a {L}\right]}{\Delta },\\
& \Sigma \dot{\phi}= \frac{a \left(a^2-\Delta +\mathcal{H}\right) \left[{E} \left(a^2+\mathcal{H}\right)-a {L}\right]}{\Delta  \Sigma  \left(a^2+\mathcal{H}-\Sigma \right)}-\frac{a \left[a^2 {E}-a {L}+{E} (\mathcal{H}-\Delta )\right]}{\Delta  \left(a^2+\mathcal{H}-\Sigma \right)},\\
& \Sigma^2 \dot{r}^2 = \mathcal{R} = \left[{E} \left(a^2+\mathcal{H}\right)-a {L}\right]^2-\Delta  \left[({L}-a {E})^2+Q\right],\\
& \Sigma^2 \dot{\theta}^2 = {\Theta}= Q+ a^2 {E}^2 {\cos \theta }^2- \frac{{\cos \theta}^2}{{\sin \theta}^2} {L}^2\label{eq161},
\end{align}
where, $Q$ is the Cater constant. As the photon sphere condition should satisfy $\dot{r}=0$ and $\ddot{r}=0$, so we have $R=0$ and ${R}'=0$. In this consideration, by introducing two impact parameters, $\xi=L/E$ and $\eta = Q/E^2$, it yields
\begin{align}\label{eq17}
& R = \left[a^2-a \xi +\mathcal{H}\right]^2 E^2-\Delta E^2 \left(a^2-2 a \xi +\eta +\xi ^2\right) =0,\\
& R'= 2 \mathcal{H}' E^2 \left[a (a-\xi )+\mathcal{H}\right]-\left(a^2-2 a \xi +\eta +\xi ^2\right) \Delta' E^2 =0.
\end{align}
By solving above equations, one can obtain
\begin{align}\label{eq18}
& \xi = \frac{a^2 \Delta'-2 \Delta \mathcal{H}'+\mathcal{H} \Delta'}{a \Delta'} |_{r=r_p},\\
& \eta = \frac{4 \Delta  \left(a^2-\Delta \right) \mathcal{H}'^2+4 \Delta  \mathcal{H} \mathcal{H}' \Delta'-\mathcal{H}^2 \Delta'^2}{a^2 \Delta'^2} |_{r=r_p}.
\end{align}
Due to the non-negativity of $\Theta$ in Eq.(\ref{eq16}), it immediately presents the condition for the photon region
\begin{align}\label{eq19}
\eta - \xi ^2 {\cot \theta} ^2 \geq -a^2 {\cos \theta}^2.
\end{align}
For each point ($r_p, \theta_p$), the null geodesic will oscillate in the $r$ or $\theta$ direction for the fixed $\theta_p$ or $r_p$.
When an observer located at position ($r_o, \theta_o$) in the Boyer-Lindquist coordinates, we introduce the orthonormal tetrad as
\begin{align}\label{eq20}
&e_0=e_{(t)}=\left(\sqrt{\frac{-g_{\phi \phi}}{g_{tt} g_{\phi \phi}-  g_{t \phi}^2} } ,0,0,  \frac{-g_{t \phi}}{ g_{\phi \phi}}  \sqrt{\frac{-g_{\phi \phi}}{g_{tt} g_{\phi \phi}-  g_{t \phi}^2} }   \right), \\
&e_1=-e_{(r)}=\left(0,\frac{-1}{\sqrt{g_{rr}}},0,0   \right),\\
&e_2=e_{(\theta)}=\left(0,0,\frac{1}{\sqrt{g_{\theta \theta}}},0   \right),\\
&e_3=-e_{(\phi)}=\left(0,0,0,\frac{-1}{\sqrt{g_{\phi \phi}}}  \right),\label{eq201}
\end{align}
where, $e_0$ is the timelike vector, which can be treated as the four-velocity of the observer, $e_1$ gives the spatial direction towards the center of the black hole, and $g_{\mu\nu}$ is the background metric of the black hole. This local rest frame is always the usual zero-angular-momentum-observer (ZAMO) tetrad.
With the aid of this tetrad, we show the illumination in Fig.\ref{fig2} to obtain the shadow and image of the black hole, where the method of stereographic projection is used.\\

The observer is located at ($r_O, \theta_O$) in the coordinates ($t,r,\theta, \phi$), which corresponding to the point $O$ in Fig.\ref{fig2}. The red line with arrow originated from $O$ represents the propagated direction of the light ray. And, the vector $\overrightarrow{OP}$ denotes the tangent vector of the null geodesic at $O$.
To further obtain the shadow and image of black hole, we take the $\overrightarrow{OP}$ as a radius and $O$ as the centre to plot a three-dimensional sphere, and place the origin of the tetrad (\ref{eq20}) at $O'$. The line $O'S$ passed through the point $O$, as the diameter of the sphere, is on a line with $e_1$. The blue rectangular plane at $O'$ is the screen for presenting the shadow and image of black hole.
By regarding the point $S$ as the reference point, the vector $\overrightarrow{OP}$ is projected to be the vector ${O'P'}$ on the screen.
To determine the photon's position from the observer's perspective, it is useful to introduce celestial coordinates($\alpha, \beta$). For the plane $O'PS$ with the green boundary of the sphere, the first celestial coordinate, $\alpha$, is defined as the angle between $OO'$ and $OP$. For the brown vertical plane of the sphere, it intersects with the front side of the plane $O'PS$ at point $Q$. The other celestial coordinate, $\beta$, is defined as the angle between $OQ$ and $e_2$.

For the null geodesic, i.e., $s(\tau) = {t(\tau), r(\tau), \theta(\tau), \phi(\tau)}$, its tangent should be a linear combination of $e_i$\footnote{Here $i=0,1,2,3.$}, which form can be expressed as
\begin{align}\label{eq21}
\dot{s} = | \overrightarrow{OP} | \left(-\chi e_0 + \cos \alpha e_1+ \sin \beta \cos \alpha e_2 + \sin \beta \sin \alpha e_3\right),
\end{align}
where, the negative sign ensures that the tangent vector is oriented towards the past, and the symbol $\cdot$ denotes the partial derivative with respect to the affine parameter $\tau$.
And, we note that the path of the photon does not depend on its energy. Therefore, we set the energy of photon in the camera's frame, $E$, to be unity, i.e., $E_{camera} = 1 = | \overrightarrow{OP} | \cdot  \chi  =- \frac{E}{ \sqrt{g_{tt}}} |_{(r_O,\theta_O)} $.
In addition, the four-momentum of photons in the ZAMO's frame can be rewrote as $p_{(\mu)} = p_{\nu} e^{\nu}_{(\mu)}$, where $e^{\nu}_{(\mu)}$ is given by Eqs.(\ref{eq20}-\ref{eq201}).
Since the values of the 4-momentum of a photon can be obtained by Eqs.(\ref{eq16}-\ref{eq161}), we can easy extract the momentum of photons $p_{(\mu)}$. By following the method in \cite{Hu:2020usx}, the connection between the celestial coordinates and the four-momentum of photons $p_{(\mu)}$ is represented as
$\cos \alpha = {p^{(1)}}/{p^{(0)}}, \tan \beta = {p^{(3)}}/{p^{(2)}}.$
Specifically, this two celestial coordinates for the rotating LQG black holes are,
\begin{align}\label{eq23}
\alpha &= \arccos \frac{\sqrt{\mathcal{M} + \mathcal{A}^2 \left(\Delta -a^2 \sin ^2\theta \right)} \sqrt{\mathcal{N} - 2 a \mathcal{H}  (1-\mathcal{F}) \xi ({r_p})-\mathcal{F} \mathcal{H} \xi ({r_p})^2-\Delta  \eta ({r_p})}}{\sqrt{\Delta } \left(a^2 \cos^2\theta+\mathcal{H}\right) \left[\mathcal{A}-a \xi ({r_p}) \left(\mathcal{H} - \mathcal{F} \mathcal{H}\right)\right]},\\
\beta &= \arctan \frac{\xi(r_p) \left(a^2 \cos^2\theta+\mathcal{H} \right)  }{  \sqrt{\mathcal{A} \sin^2 \theta}  \sqrt{ \left[a^2 \cos^2\theta+\eta(r_p)-\cot^2\theta \xi(r_p)^2\right]}},
\end{align}
where, $\mathcal{N} = \mathcal{A}-a^2 \Delta  \cos^2\theta$, $\mathcal{M} = \mathcal{A}  a^2 \sin ^2\theta (\mathcal{H}- \mathcal{F} \mathcal{H})^2 $.
With the help of Eq.(\ref{eq22}), we introduce the Cartesian coordinates system ($x,y$) on the screen, it yields
\begin{align}\label{eq22}
x = -2 \tan \frac{\alpha}{2} \sin \beta, \quad y = - 2 \tan \frac{\alpha}{2} \cos \beta.
\end{align}
Therefore, one can present the boundary of the shadow of the rotating black holes on the screen. And more importantly, we will discuss the effects of quantum parameter $\lambda$, rotating parameter $a$, inclination angle $\theta_o$ and  observed distance $r_o$ on the black hole shadow.

\begin{figure}[!h]
\makeatletter
\renewcommand{\@thesubfigure}{\hskip\subfiglabelskip}
\makeatother
\centering % \begin{center}/\end{center} takes some additional vertical space
\subfigure[$ $]{
\setcounter{subfigure}{0}
\subfigure[$(a) r_{o}=100, \theta_{o}=\pi/2$]{\includegraphics[width=.3\textwidth]{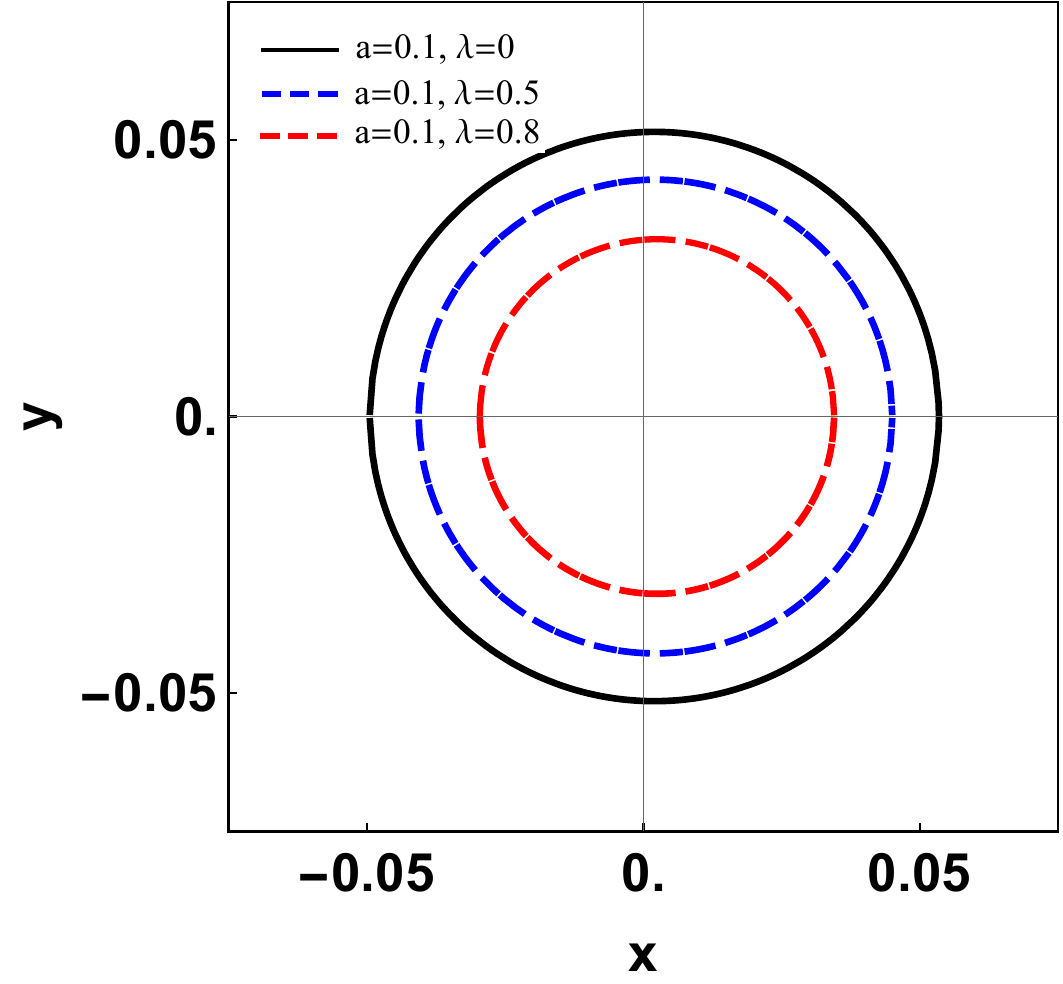}}
\hspace{1cm}
\subfigure[$(b) r_{o}=100, \theta_{o}=\pi/2$]{\includegraphics[width=.3\textwidth]{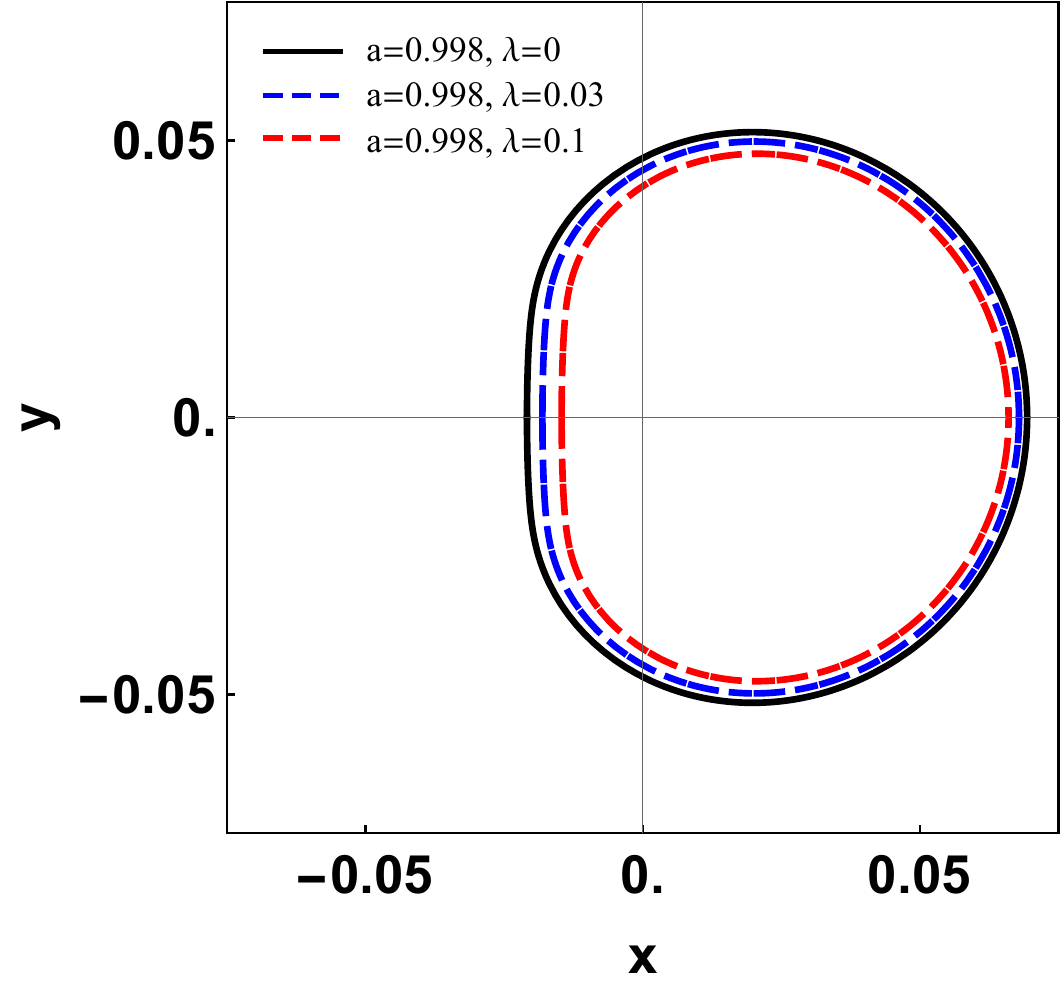}}
}
\caption{\label{fig3}  The shadow versus quantum parameter $\lambda$. }
\end{figure}

In Fig.\ref{fig3}, it shows the influence of $\lambda$ on the black hole shadow for different values of $a$. Clearly, one can see that the parameter $\lambda$ always decreases the size of shadow for both $a=0.1$ and $a=0.998$, and seems to have no effects on the deformation of shadow. According to the Fig.\ref{fig1}, we know that the acceptable range of values for $\lambda$ is bigger and bigger with the decrease of the rotating parameter $a$. So, when $a=0.1$, the broad range of $\lambda$ leads to substantial variations in the shadow's size. Conversely, with $a=0.998$, the narrower range of $\lambda$ results in a smaller changes in the shadow size.

\vspace{-0.1cm}
\begin{figure}[!h]
\makeatletter
\renewcommand{\@thesubfigure}{\hskip\subfiglabelskip}
\makeatother
\centering % \begin{center}/\end{center} takes some additional vertical space
\subfigure[$ $]{
\setcounter{subfigure}{0}
\subfigure[$(a) r_{o}=100, \theta_{o}=\pi/2$]{\includegraphics[width=.3\textwidth]{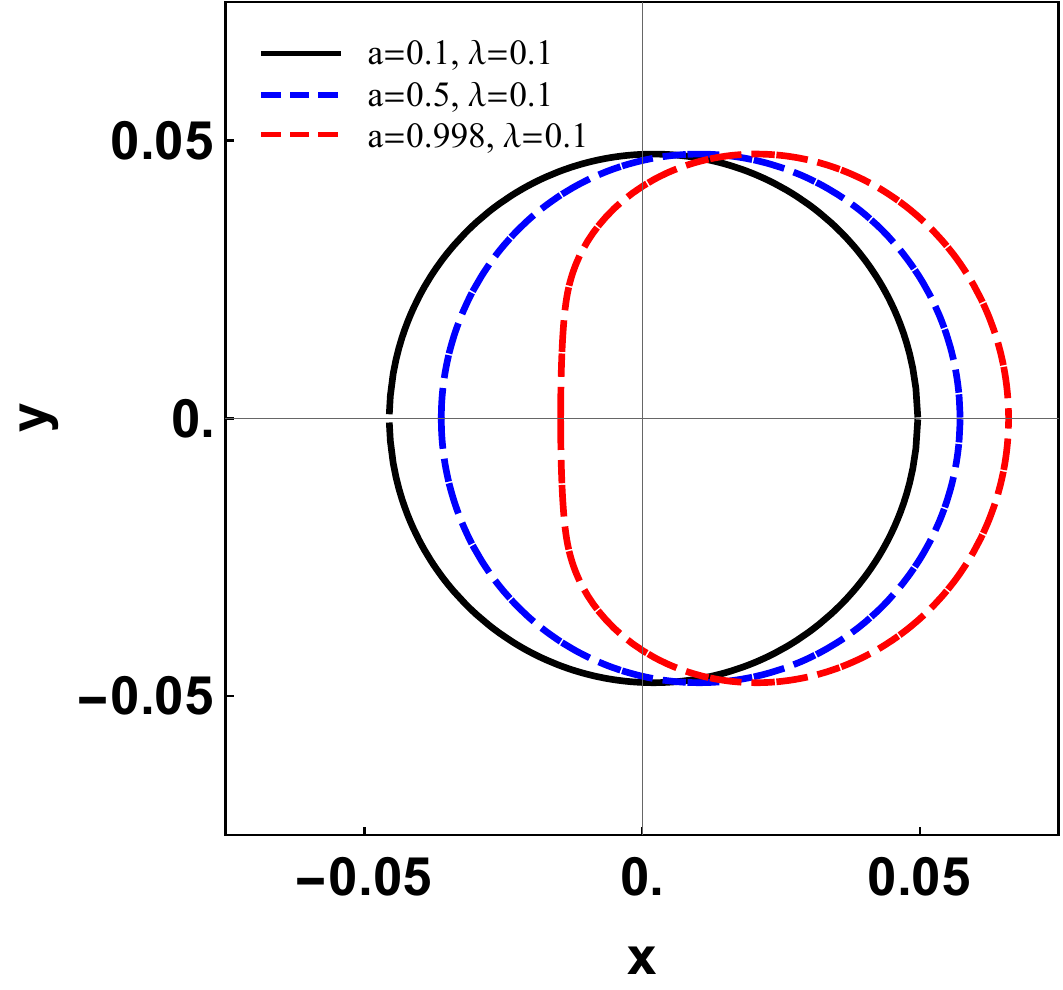}}
\subfigure[$(b) a=0.998, r_{o}=100$]{\includegraphics[width=.3\textwidth]{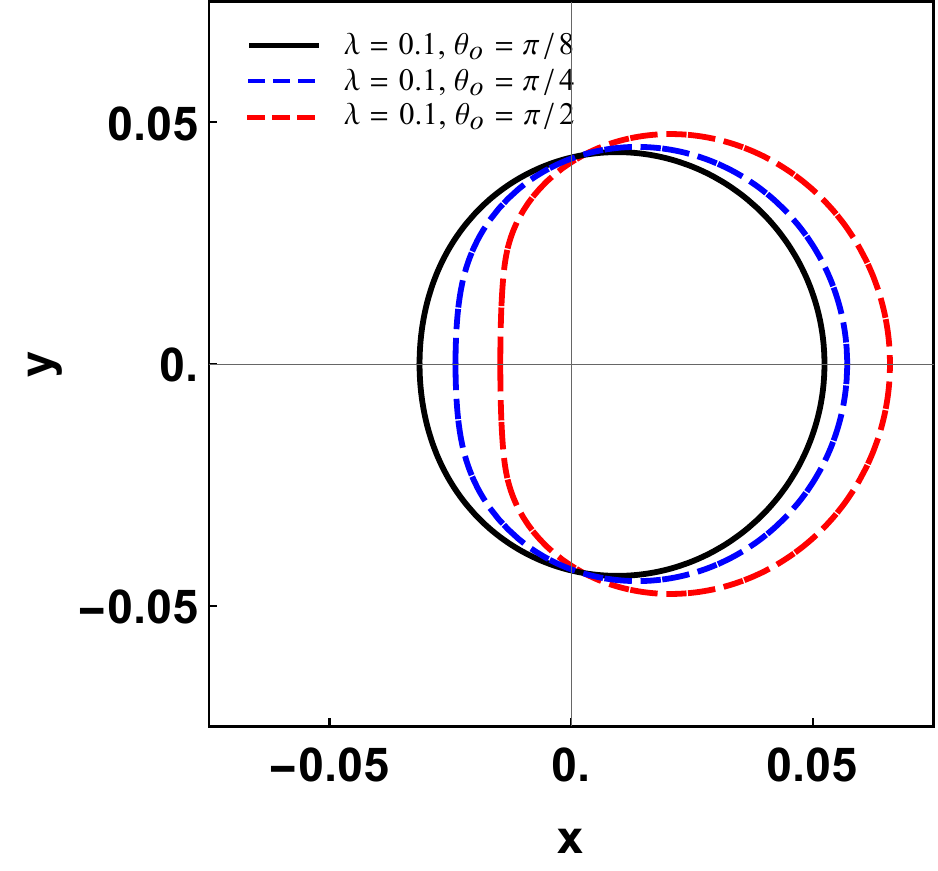}}
\subfigure[$(c) a=0.998, \theta_{o}=\pi/2$]{\includegraphics[width=.3\textwidth]{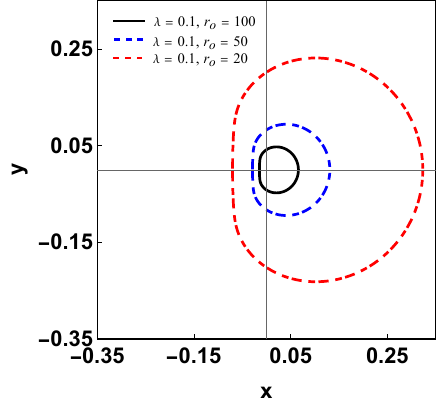}}
}
\caption{\label{fig4}  The shadow versus parameters $a, r_o,$ and $\theta_o$. }
\end{figure}

In Fig.\ref{fig4}, we fixed $\lambda = 0.1$, and varied other corresponding parameters, i.e., $a, r_o, \theta_o$. When the observer located at the equatorial plane ($\theta_o = \pi/2$), the deformation of shadow is more and more obvious with the increase of $a$ (Fig.\ref{fig4}(a)), and the size of shadow increased with the decrease of the observed distance (Fig.\ref{fig4}(c)). Furthermore, the increase of the viewing angle of the observer also transform the circular shadow into a ``D'' shape (Fig.\ref{fig4}(b)).

Next, we continue to focus on the deviation from the circularity $\delta_s$ and the size $R_s$ of the shadow cast by the rotating black holes, which defined by Hioki and Maeda \cite{Hioki:2009na}.

\begin{figure}[!h]
    \centering
\includegraphics[width=0.4\linewidth]{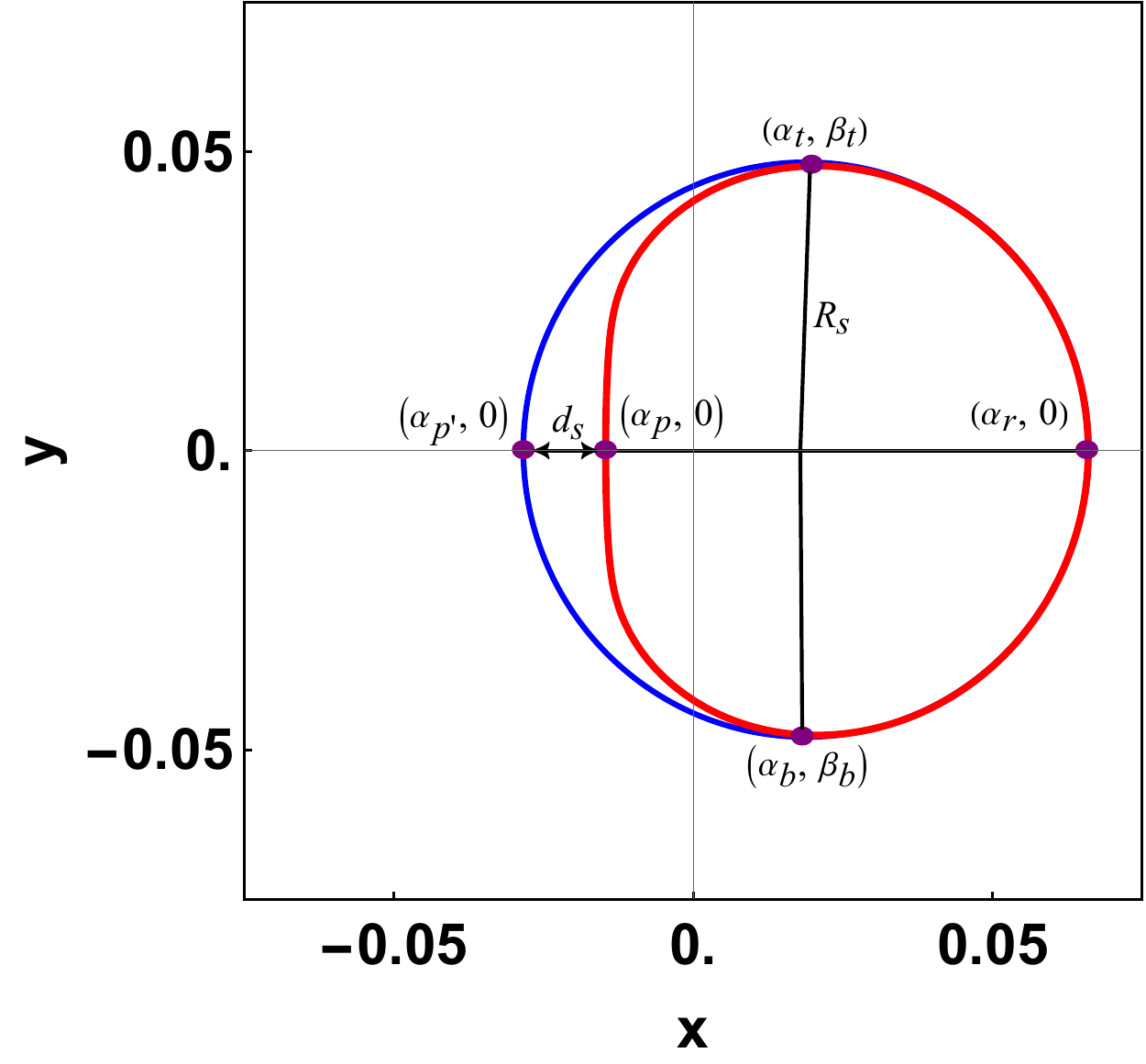}
    \caption{Black hole shadow and reference circle.}
    \label{fig5}
\end{figure}

In Fig.\ref{fig5}, the five reference points ($\alpha_t, \beta_t$), ($\alpha_b, \beta_b$), ($\alpha_r, 0$), ($\alpha_p, 0$) and ($\alpha_{p'}, 0$) correspond to the top, bottom, rightmost, and leftmost points of the shadow, leftmost point of the reference circle, respectively. The deviation from the circularity ($\delta_s$) and the size ($R_s$) are defined as
\begin{align}\label{eq24}
R_s = \frac{(\alpha_t - \alpha_r)^2 + \beta_t^2}{2 | \alpha_t - \alpha_r|}, \quad
\delta_s= \frac{ |\alpha_{p'}- \alpha_p|}{R_s}.
\end{align}
In Tables 1 and 2, we intuitively show the deviation from the circularity $\delta_s$ and the size $R_s$ of the shadow cast by the non-singular rotating black holes.

\begin{center}
{\footnotesize{\bf Table 1.} The observable $R_s$  and $\delta_s$ for $a=0.9$.\\
\vspace{1mm}
\label{tab2}
\begin{tabular}{ccccccccccc}
\hline       &{$\lambda=0.01$}  &{$\lambda=0.04$}  &{$\lambda=0.07$} &{$\lambda=0.1$} &{$\lambda=0.13$} &{$\lambda=0.16$} &{$\lambda=0.19$} &{$\lambda=0.22$} &{$\lambda=0.25$}     \\ \hline
{$ R_s$}  &{0.05067}  &{0.04941}  &{0.04844} &{0.04758} &{0.04679} &{0.04604} &{0.04530} &{0.04462} &{0.04394}        \\
{$\delta_s$} &{0.14392}  &{0.15415}  &{0.16230}  &{0.16960} &{0.17650} &{0.18315}  &{0.18955}  &{0.19582} &{0.20197}    \\
\hline
\end{tabular}}
\end{center}

\begin{center}
{\footnotesize{\bf Table 2.} The observable $R_s$  and $\delta_s$ for $\lambda=0.1$.\\
\label{tab1}
\begin{tabular}{ccccccccccc}
\hline   &{$a=0.1$}  &{$a=0.2$}  &{$a=0.3$} &{$a=0.4$} &{$a=0.5$} &{$a=0.6$} &{$a=0.7$} &{$a=0.8$} &{$a=0.9$}     \\ \hline
{$ R_s$}      &{0.047579}  &{0.0475791}  &{0.0475794} &{0.0475797} &{0.0475802} &{0.0475807} &{0.0475813} &{0.0475821} &{0.0475829}      \\
{$ \delta_s$} &{0.001366}  &{0.005500}  &{0.0126011} &{0.0229739} &{0.0371891} &{0.0561589} &{0.0815062} &{0.116492} &{0.1696300}      \\
\hline
\end{tabular}}
\end{center}

It turns out from Tables 1 and 2 that the shadow's size $R_s$ and the deviation from the circularity $\delta_s$ are all increased with the rotating parameter $a$ for the fixed parameter $\lambda$.
When $a$ is held constant, as $\lambda$ increases, the size of the shadow gradually decrease, while the deviation from circularity progressively increases.
Obviously, the trend of $R_s$ with $\lambda$ can actually be observed in Fig.\ref{fig3}(b). However, the changes in circularity deviation for $\lambda$ are very subtle and difficult to discern from Fig.\ref{fig3}(b). The effect of $\lambda$ on the circularity deviation can be observed by Table 1.

In addition, we can also approximately estimate the angular radius of the rotating black holes, and further present the effect of $\lambda$ on it. In general, the angular radius, $\Theta_{BH}$, is defined as $\Theta_{BH} = \tilde{R}_{s}\frac{\mathcal{M}}{D_O}$, where $D_O$ is the distance between black hole and observer, and the radius $\tilde{R}_s$ with its screen located at the position of black hole is related to the screen's shadow radius $R_s$ obtained in Eq.(\ref{eq23}), which can be obtained by using the simple geometrical relationship. As described by \cite{Amarilla:2011fx,Liu:2024lbi,Liu:2024soc}, when a black hole with mass $\mathcal{M}$ is located far from the observer, the angular radius $\Theta_{BH}$ observed can be quantitatively described as $\Theta_{BH} = 9.87098 \tilde{R}_s \left(\frac{\mathcal{M}}{ M_{\astrosun}}\right) \left( \frac{1 kpc}{ D_O}  \right) \mu$$as$. Using this form, we take the SgrA* and M87* as examples, to calculate the angular radius of the black hole in the background of Eq.(\ref{eq7}). For the $M87^*$, the distance from the Earth is $D_O = 16.8Mpc$, and the estimated mass of black hole is $\mathcal{M} = (6.5 \pm 0.7) \times 10^6 { M_{\astrosun}}$. And, the the actual shadow diameter should be $\Theta_{M87^*} = (37.8 \pm 2.7)$$\mu$$as$, which is a $10\%$ discrepancy between the image and shadow diameters \cite{Capozziello:2023tbo}. For the $SgrA^*$, the observer distance, the estimated mass of black hole, and the shadow diameter are  $D_O = 8kpc$,  $\mathcal{M} = (4.0_{-0.6}^{+1.1}) \times 10^6 { M_{\astrosun}}$ and $\Theta_{SgrA^*} = (48.7 \pm7)$$\mu$$as$ \cite{KumarWalia:2022aop}, respectively.

\vspace{-0.1cm}
\begin{figure}[!h]
\makeatletter
\renewcommand{\@thesubfigure}{\hskip\subfiglabelskip}
\makeatother
\centering
\subfigure[$ $]{
\setcounter{subfigure}{0}
\subfigure[$(a)$]{\includegraphics[width=.4\textwidth]{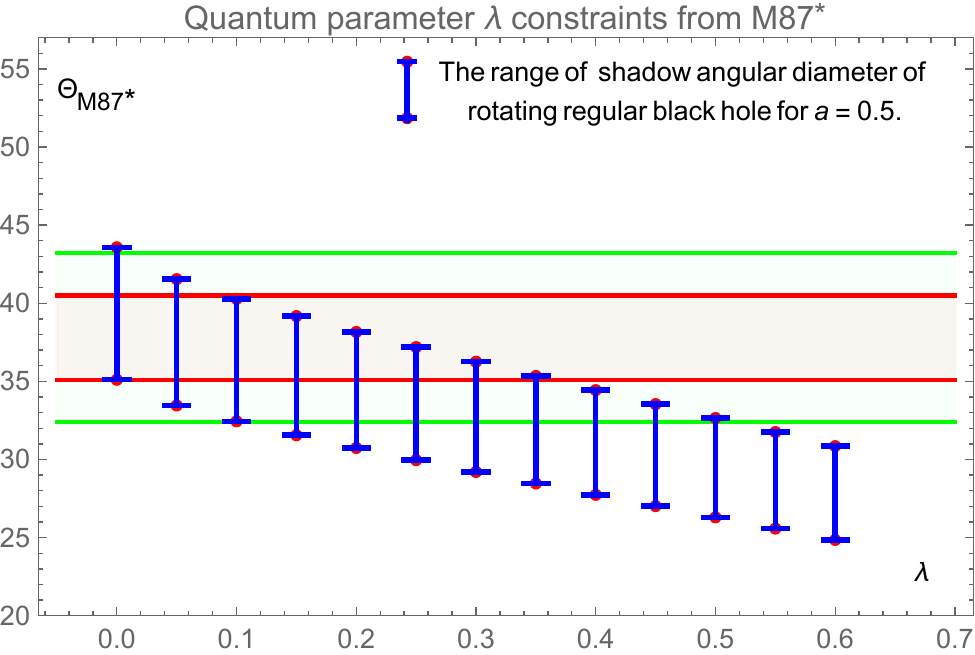}}
\hspace{1cm}
\subfigure[$(b)$]{\includegraphics[width=.4\textwidth]{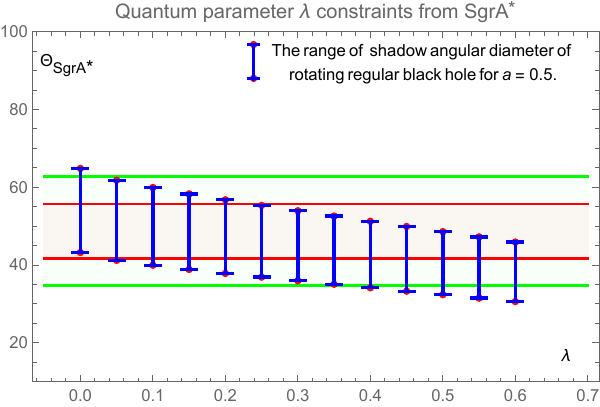}}
}
\caption{\label{fig6} The shadow angular diameter of non-singular rotating black hole for $a = 0.5$. }
\end{figure}
In Fig.\ref{fig6}, the range between two green lines denotes the 2$\sigma$ of the shadow angular diameter $\Theta_{BH}$, while the range between two red lines represents the 1$\sigma$ of the shadow angular diameter $\Theta_{BH}$.
It is true that the maximum value of $\lambda$ for the rotating black hole (\ref{eq7}) can be up to 0.62 from the Fig.\ref{fig1}, when $a=0.5$.
For $M87^*$, the observed angular diameter decreases as the quantum parameter $\lambda$ increases. When $\lambda=0$, the results correspond to the case of the Kerr black hole. However, for $\lambda \simeq 0.37$ and $\lambda \simeq 0.52$, the observed angular diameter of the black hole falls outside the $1\sigma$ and $2\sigma$ observational bounds of $M87^*$'s angular diameter, respectively. In contrast, for $SgrA^*$, we find that the entire range of permissible $\lambda$ values (from 0 to 0.62) lies within the $1\sigma$ observational bounds of $SgrA^*$'s angular diameter. This indicates that the astronomical observations of M87* impose significantly stronger constraints on the observed properties of rotating black holes compared to $SgrA^*$.

\section{Shadows illuminated by the celestial light source }\label{sec4}

In this section, we will employ the backward ray-tracing method to characterize the image of the rotating black hole in the framework of a celestial light source. For the celestial sphere light source model, the black hole is located at the center of the celestial sphere, with its rotation axis pointing towards the North Pole, and is also considerably smaller than both the celestial sphere and the distance between the observer and the origin. The observer is assumed to be situated in the equatorial plane within the sphere. And, the celestial sphere is marked with four into four quadrants, which correspond to four distinct color designations: the green quadrants ($0 \leq \theta \leq \pi/2$ and $0 \leq \phi \leq \pi$), the red quadrants ($0 \leq \theta \leq \pi/2$ and $\pi \leq \phi \leq 2\pi$), the blue quadrants ($\pi/2 \leq \theta \leq \pi$ and $0 \leq \phi \leq \pi$), and the yellow quadrants ($\pi/2 \leq \theta \leq \pi$ and $\pi \leq \phi \leq 2\pi$).
The backward ray-tracing method allows one to trace fewer light rays, without concerning those emitted from the light source that do not reach to the observer. It provides a more convenient way to study black hole images. We follow the numerical strategy outlined in \cite{Hu:2020usx}, which consists of two parts: the first is a fisheye camera model, primarily used for stereographically projecting the momentum of photons $p_{\mu}$ onto the screen; the second is the integration of the equations of motion, where the photon's motion equations are integrated backward along the null geodesics from the observer.

The equations of motion from the Hamiltonian formulation can be expressed as
\begin{align}\label{eq25}
\dot x^{\mu} = \frac{\partial H }{\partial p_{\mu}}, \quad \dot p_{\mu} = - \frac{\partial H}{\partial x^{\mu}},
\end{align}
where the overdot implies differentiation with respect to the affine parameter $\tau$.
By using the backward ray-tracing method, once the position of the light ray reaching the celestial sphere is determined, its corresponding color is also identified, where the light rays that reach the event horizon being marked in black.
In the background of the metric (\ref{eq7}), we first solve the above equation (\ref{eq25}). By employing the fisheye camera model, then we can successfully obtain the shadow region of the black hole on the screen under the celestial sphere light source, as well as the distortion behavior of light rays around the black hole.

\vspace{-0.1cm}
\begin{figure}[!h]
\makeatletter
\renewcommand{\@thesubfigure}{\hskip\subfiglabelskip}
\makeatother
\centering
\subfigure[$ $]{
\setcounter{subfigure}{0}
\subfigure[$(a):r_O=100, \text{fov}={\pi}/{10}, a=0.1, \lambda =0.01$]{\includegraphics[width=.33\textwidth]{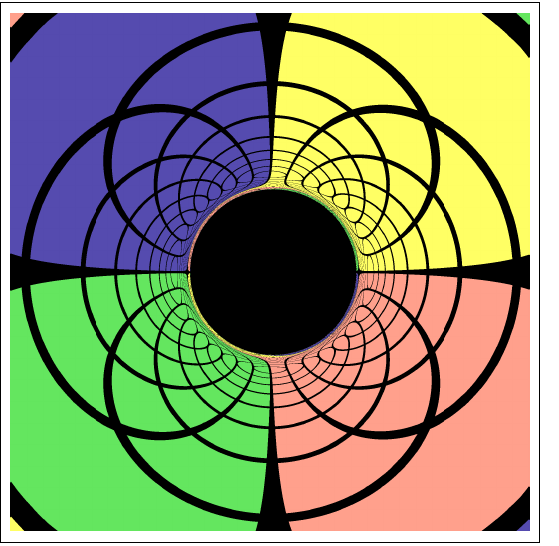}}
\hspace{1cm}
\subfigure[$(b):r_O=100, \text{fov}={\pi}/{10}, a=0.1, \lambda =0.9$]{\includegraphics[width=.33\textwidth]{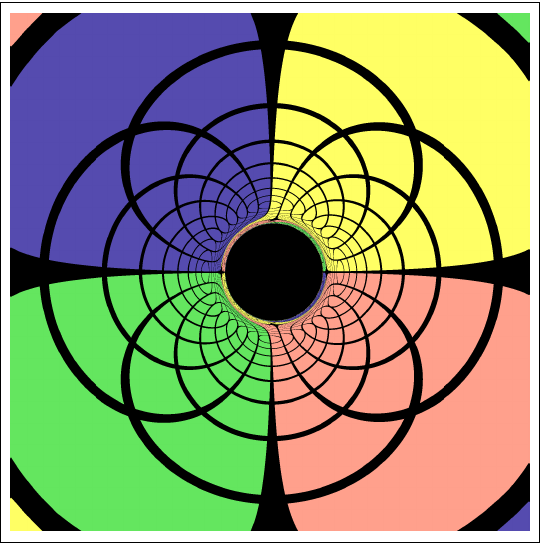}}
}
\subfigure[$ $]{
\setcounter{subfigure}{2}
\subfigure[$(c): r_O=100, \text{fov}={\pi}/{10}, a=0.998, \lambda =0.01$]{\includegraphics[width=.33\textwidth]{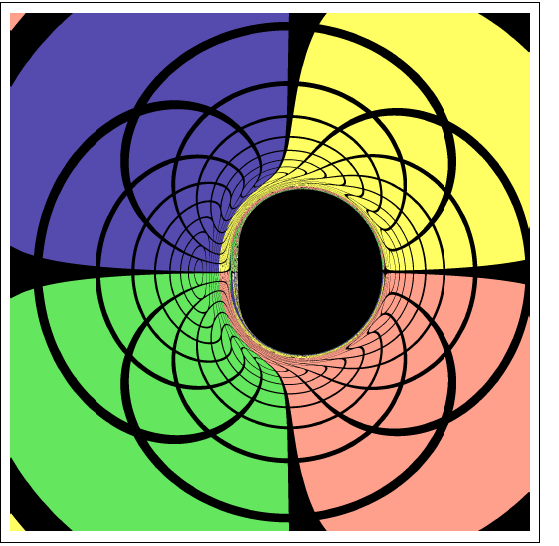}}
\hspace{1cm}
\subfigure[$(d): r_O=100, \text{fov}={\pi}/{10}, a=0.998, \lambda =0.1 $]{\includegraphics[width=.33\textwidth]{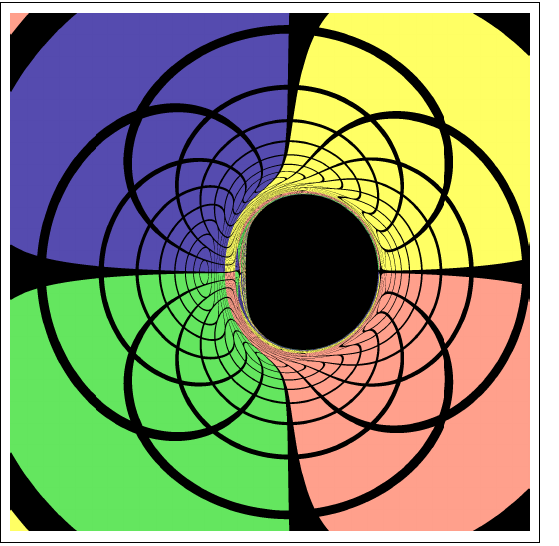}}
}
\caption{\label{fig7} The shadow regions of non-singular rotating black hole. }
\end{figure}

The Fig.\ref{fig1} shows that the maximum values of the non-singular  black hole parameter $\lambda$ are approximately $0.923$ and $0.111$ when $a=0.1$ and $a=0.998$, respectively. Based on the parameter constraints, we have presented the black hole shadow regions under the four-color celestial sphere light source, where the symbol $fov$ represents the field of view of the camera, and the observer is positioned at
$(r_o, \theta_o)$. As shown in Fig.\ref{fig7}, when $a=0.1$, the shape of the black hole shadow is essentially circular, and its size decreases with the increase of $\lambda$. When $a=0.998$, the shadow takes on a ``D" shape, and its size also decrease with the increase $\lambda$. It is important to note that, compared to the case when $a=0.1$, the range of variation for parameter $\lambda$ is smaller when $a=0.998$, resulting in a more limited reduction in the shadow area. Additionally, we observe that the impact of parameter $\lambda$ on the shadow region and the surrounding image appears to be quite regular, without any unusual changes such as the  cuspy shadow observed in \cite{Cunha:2017eoe}.\\

\vspace{-0.1cm}
\begin{figure}[!h]
\makeatletter
\renewcommand{\@thesubfigure}{\hskip\subfiglabelskip}
\makeatother
\centering
\subfigure[$ $]{
\setcounter{subfigure}{0}
\subfigure[$(a)$: $\lambda=0$]{\includegraphics[width=.33\textwidth]{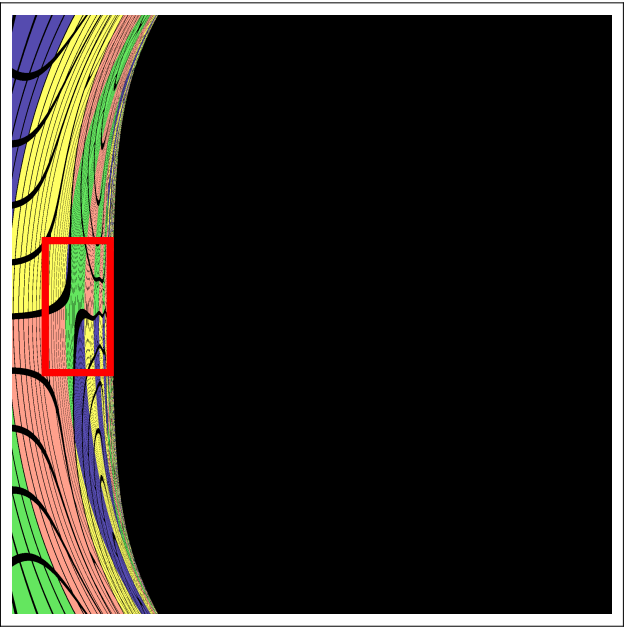}}
\hspace{1cm}
\subfigure[$(b)$: $\lambda=0.1$]{\includegraphics[width=.33\textwidth]{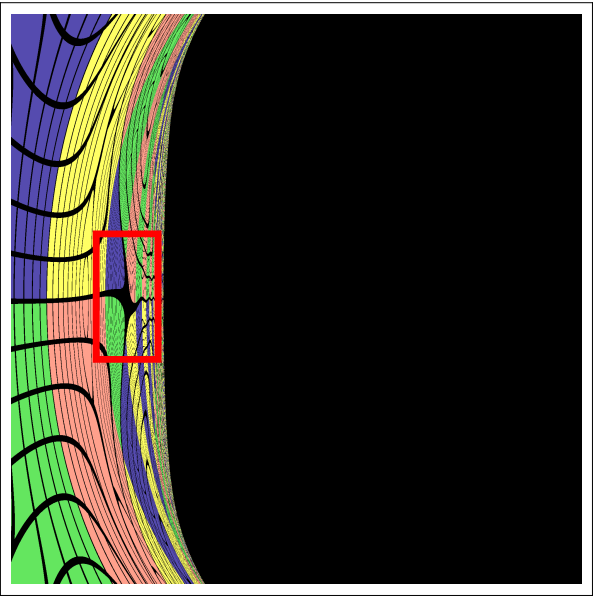}}
}
\caption{\label{fig88} The shadow regions of the non-singular black hole.}
\end{figure}
%\begin{tikzpicture}[remember %picture, overlay]
    % 矩形坐标 (x0, y0) - (x1, y1)
    %\draw[red, thick] (2, 2) %rectangle (5, 5);
%\end{tikzpicture}

The differences observed in Fig.\ref{fig88} are subtle and can only be discerned upon magnifying the image and closely examining it. Focus on the rectangular area in Fig.\ref{fig88}, we can see that the black latitude and longitude lines do not intersect for the Kerr metric($\lambda=0$), while in the non-singular black hole, they converge at a certain point. Additionally, in the Kerr black hole, the blue region within the red box predominantly appears in the lower half of the box, whereas in the non-singular black hole, a noticeable larger blue area appears in the upper half of the box. We believe that due to the very weak manifestation of QG effects in non-singular black holes, the impact on the black hole shadow is also minimal.\\

\section{Images illuminated by the thin accretion disk}\label{sec5}
\begin{figure}[!h]
    \centering
\includegraphics[width=0.55\linewidth]{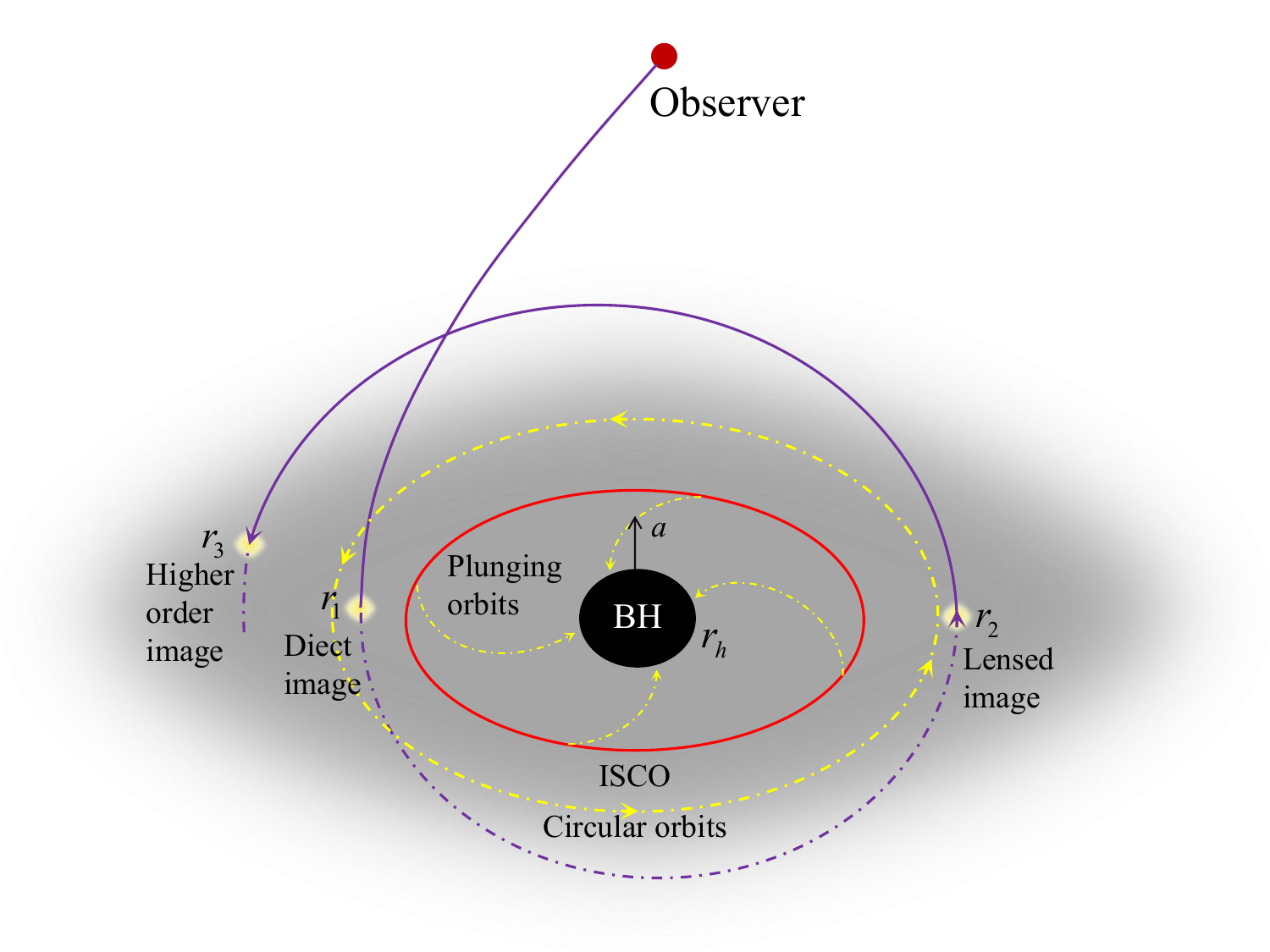}
    \caption{Imaging the black hole with a thin accretion disks.}
    \label{fig41}
\end{figure}

Typically, the millimeter-wave images of supermassive black holes are dominated by their surrounding accretion disks. In view of this, we use the accretion disk model from reference \cite{Hou:2022eev} as the light source to further study the imaging characteristics of non-singular black holes with a thin accretion  disk in this section. For simplicity, we assume that the accretion disk is geometrically and optically thin, situated on the equatorial plane, while the observer is located at a distant position.
The accretion disk is divided into two parts by the innermost stable circular orbit (ISCO): the region inside the ISCO, where the accretion disk undergoes plunging motion, and the region outside the ISCO, where the accretion disk follows stable circular orbits. On the other hand, considering the suggestion from reference \cite{Hou:2022eev}, light rays may intersect with the accretion disk once, twice, three times, or even more. Each intersection contributes to the luminosity of the black hole image. We denote the position of the first intersection as $r_1$, which corresponds to the direct image of the black hole observed. The second and third intersections correspond to the lensing image and higher-order images, respectively. And, in Fig.\ref{fig41}, we present a schematic diagram of the accretion disk images.

In this paper, we assume that the accretion disk starts from the event horizon of the black hole ($r_h$) and extends to a very distant point ($r_f$), with $r_f$ is set to be 10000. And, the position of the observer ($r_O$) is located in the region $r_+ \ll r_o < r_f$.
To obtain the image of the black hole in the background of this thin disk, we first need to find the location of the ISCO. Generally, the position of the ISCO is determined by the following conditions
\begin{align}\label{eq26}
V_{eff}|_{r=r_{ISCO}} &= 0, \\
\partial_r V_{eff}|_{r=r_{ISCO}} &= 0, \\
\partial_r^2 V_{eff}|_{r=r_{ISCO}} &= 0,
\end{align}
where, the sign $ V_{eff}$ is the effective potential function. For a massive neutral particle with four-velocity $u^a$, its form is
\begin{align}\label{eq27}
V_{eff}(r,\tilde{E},\tilde{L})=(1+g^{tt}\tilde{E}^2 + g^{tt}\tilde{L}^2 - 2g^{t\phi}\tilde{E}\tilde{L})|_{\theta=\pi/2},
\end{align}
where two constants of motion, $\tilde{E}$ and $\tilde{L}$ as the conserved quantities which represent the specific energy and the specific angular momentum of massive neutral particle. And, they are
\begin{align}\label{eq28}
\tilde{E}=-\frac{g_{tt} + g_{t\phi} \Omega}{\sqrt{-g_{tt} - 2 g_{t\phi} \Omega -g_{\phi \phi} \Omega^2}}, \qquad
\tilde{L}=\frac{g_{t\phi} + g_{\phi\phi} \Omega}{\sqrt{-g_{tt} - 2 g_{t\phi} \Omega -g_{\phi \phi} \Omega^2}},
\end{align}
with
\begin{align}\label{eq29}
\Omega=\frac{d\phi}{dt} = \frac{\partial_r g_{t\phi} + \sqrt{\partial_r^2 g_{t\phi} - \partial_r g_{tt} \partial_{r}g_{\phi \phi}}}{\partial_r g_{\phi \phi}}.
\end{align}
At the location of the ISCO, the conserved quantities reads $\tilde{E}_{ISCO}$ and $\tilde{L}_{ISCO}$. Outside the ISCO, the accretion flows in the accretion disk move along nearly circular orbits, and their four-velocity is given by
\begin{align}\label{eq30}
u^{\mu}_{out}= \sqrt{\frac{1}{-g_{tt} - 2 g_{t\phi} \Omega -g_{\phi \phi} \Omega^2}} \left(1,0,0,\Omega \right)_{\mid_{\theta=\pi/2}}.
\end{align}
Inside the ISCO, the accretion fluid falls from the ISCO towards the event horizon of the black hole. For convenience, we assume that the conserved quantities $\tilde{E}_{ISCO}$ and $\tilde{L}_{ISCO}$ are equal to the values at the ISCO. In this case, the four-velocity is given by
\begin{align}\label{eq31}
&u^{t}_{in}=\left(-g^{tt}\tilde{E}_{ISCO} +g^{t\phi}\tilde{L}_{ISCO}  \right)_{\mid_{\theta=\pi/2}}, \nonumber \\
&u^{r}_{in}= - \sqrt{-\frac{g_{tt} u^t_{in}u^t_{in} + 2g_{t\phi}u^t_{in}u^{\phi}_{in}+g_{\phi\phi}u^{\phi}_{in}u^{\phi}_{in} +1  }{g_{rr}}}_{\mid_{\theta=\pi/2}},\nonumber  \\
&u^{\theta}_{in}= 0, \nonumber  \\
& u^{\phi}_{in}=\left(-g^{t\phi}\tilde{E}_{ISCO} +g^{\phi\phi}\tilde{L}_{ISCO}  \right)_{\mid_{\theta=\pi/2}}.
\end{align}
In above equation, the negative sign in front of the square root indicates the direction towards the event horizon, and the range of $r$ is $r_+< r < r_{ISCO}$.

As previously mentioned, when we trace the light rays backward from the observer's position, the rays may intersect the equatorial plane's accretion disk once ($n=1$), twice ($n=2$), or even more times ($n>2$). Each intersection allows the observer to receive additional luminosity, with the intersection position denoted as $r_n$, which we define as the transfer function. Thus, by neglecting reflection effects and the thickness of the accretion disk, the intensity observed on the observer's screen can be determined as follows
\begin{align}\label{eq32}
I_{\nu_o}= \sum_{n=1}^{N} f_n g_n^3 J_n(r),
\end{align}
where $N$ is the maximum number of intersections between the light rays and the accretion disk, and $f_n$ is the fudge factor which is fixed to 1 in this paper .
And, by considering that the observational wavelength of the black hole images taken by the EHT is $1.3 mm$ ($230 GHz$), we choose the emissivity of the thin disk as a second-order polynomial in log-space
\begin{align}\label{eq33}
J(r)= exp \left[{-\frac{1}{2}z^2 -2z }\right], \quad z= \log \frac{r}{r_+}.
\end{align}
On the other hand, the redshift factor is $g_n = \nu_o/\nu_n$ with $\nu_o$ is the observed frequence on the screen, and $\nu_n$ denotes the frequence observed by the local rest frames comoving with the accretion disk. Since the accretion disk is an electrically neutral plasma and moves along timelike geodesics with $\tilde{E}$ and $\tilde{L}$, the redshift factor of these circularly orbiting fluids outside the ISCO can be written as follows
\begin{align}\label{eq34}
g_n^{out}= \frac{\xi \left(1- \gamma \frac{  p_{\phi}}{p_{t}}\right)}{\zeta \left( 1 + \Omega \frac{  p_{\phi}}{p_t} \right)} {|_{r=r_n}}, \quad r>r_{ISCO},
\end{align}
with $\gamma=\frac{g_{t\phi}}{g_{\phi\phi}}$, $\xi=\sqrt{\frac{-g_{\phi\phi}}{ g_{tt} g_{\phi\phi} -g_{t\phi}^2} }$, $\zeta = \sqrt{ \frac{-1}{g_{tt} +2 g_{t\phi} \Omega +g_{\phi\phi}\Omega^2 }}$, and $e =\frac{p_{(t)}}{p_t} =  \xi \left(1- \gamma \frac{  p_{\phi}}{p_{t}}\right) $ is the ratio of the observed energy on the screen to the energy along a null geodesic, which is fixed as $e=1$ for the asymptotically flat spacetimes since the observer located at infinity.  Within the ISCO, the accretion flow moves along critical plunging orbits with $\tilde{E}$ and $\tilde{L}$, and its radial velocity is $u^r_{in}$. In this case, the redshift factor should be expressed as
\begin{align}\label{eq35}
g_n^{in}=\frac{1}{u^r_{in} p_r/p_t - \tilde{E}_{ISCO}(g^{tt} -g^{t\phi} p_\phi/p_t ) +\tilde{L}_{ISCO}(g^{\phi \phi} p_\phi/p_t +g^{t\phi})} {|_{r=r_n}},  \quad r<r_{ISCO}.
\end{align}
Therefore, by employing the redshift factor and the emission model of the disk, we can obtain the observable image of the non-singular black hole under thin disk accretion with prograde flows in Fig.\ref{figbp} by choosing the suitable color-function for the visual quality.

\begin{figure}[!h]
\makeatletter
\renewcommand{\@thesubfigure}{\hskip\subfiglabelskip}
\makeatother
\centering % \begin{center}/\end{center} takes some additional vertical space
\subfigure[(a) The position of observer $\theta_{obs}=80^\circ$.]{
\setcounter{subfigure}{0}
\subfigure[$a=0.1,\lambda=0.09$]{\includegraphics[width=.215\textwidth]{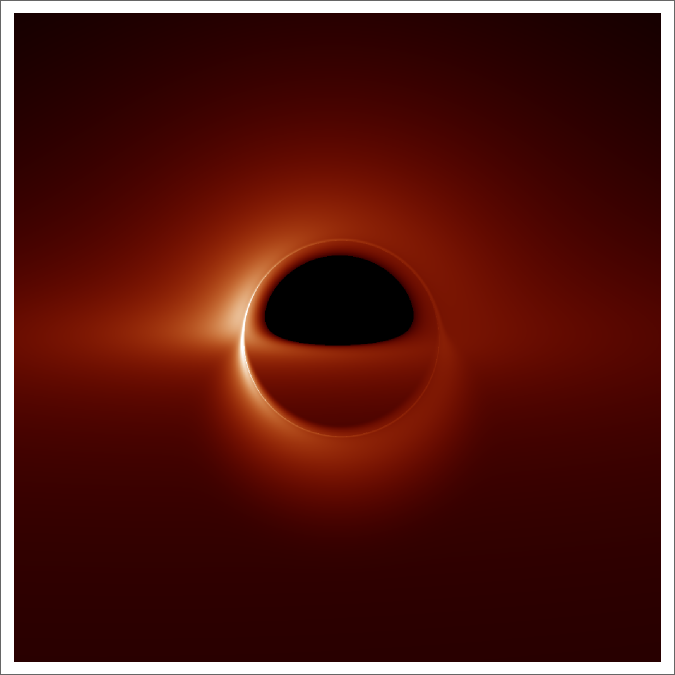}}
\subfigure[]{\includegraphics[width=.03\textwidth]{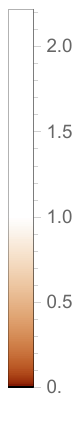}}
\subfigure[$a=0.1,\lambda=0.9$]{\includegraphics[width=.215\textwidth]{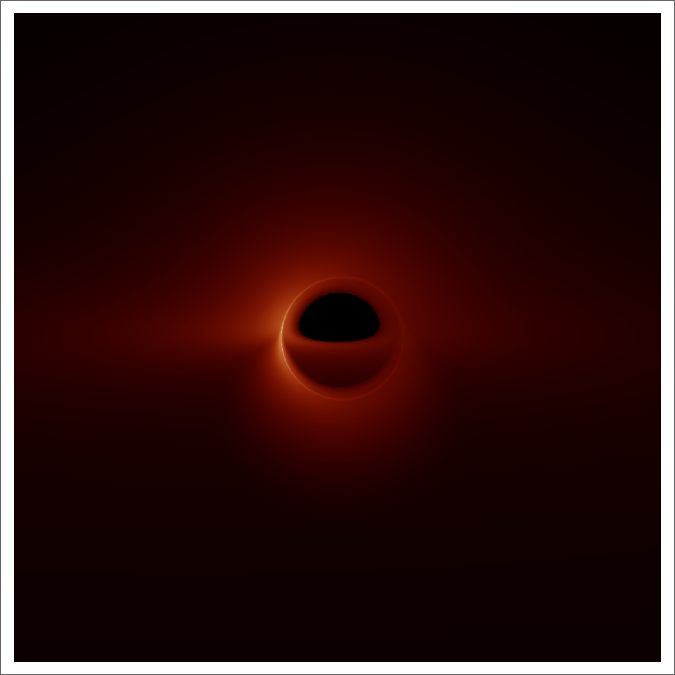}}
\subfigure[]{\includegraphics[width=.03\textwidth]{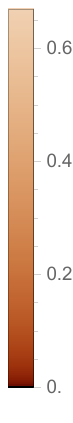}}
\subfigure[$a=0.998,\lambda=0.01$]{\includegraphics[width=.215\textwidth]{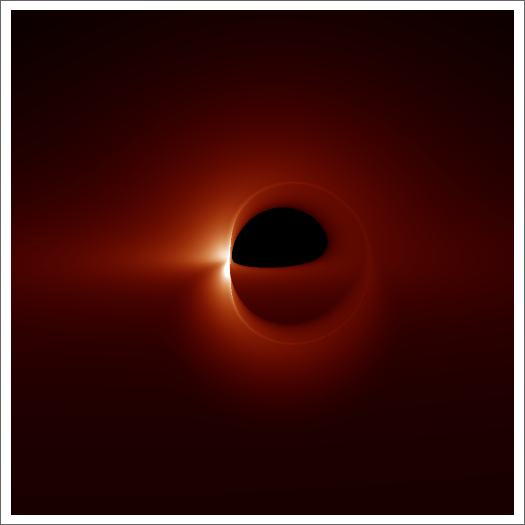}}
\subfigure[]{\includegraphics[width=.025\textwidth]{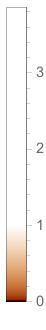}}
\subfigure[$a=0.998,\lambda=0.1$]{\includegraphics[width=.215\textwidth]{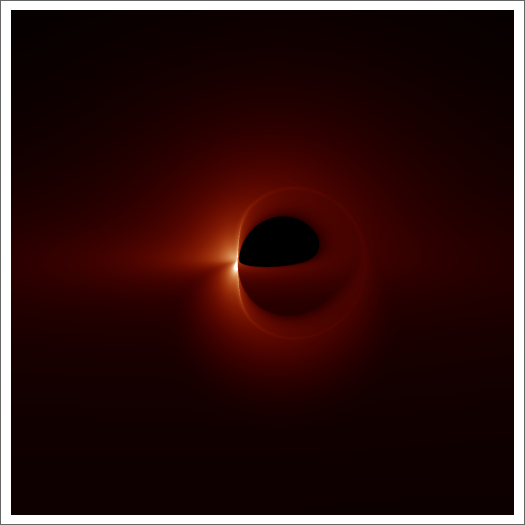}}
\subfigure[]{\includegraphics[width=.025\textwidth]{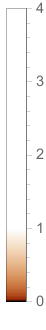}}}

\subfigure[(b) The position of observer $\theta_{obs}=163^\circ$.]{
\setcounter{subfigure}{0}
\subfigure[$a=0.1,\lambda=0.09$]{\includegraphics[width=.21\textwidth]{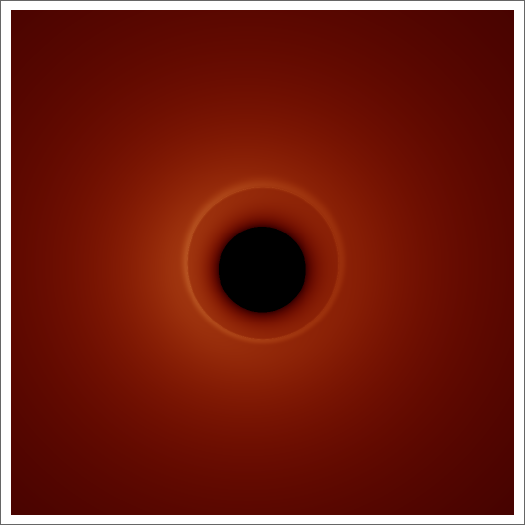}}
\subfigure[]{\includegraphics[width=.035\textwidth]{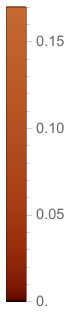}}
\subfigure[$a=0.1,\lambda=0.9$]{\includegraphics[width=.21\textwidth]{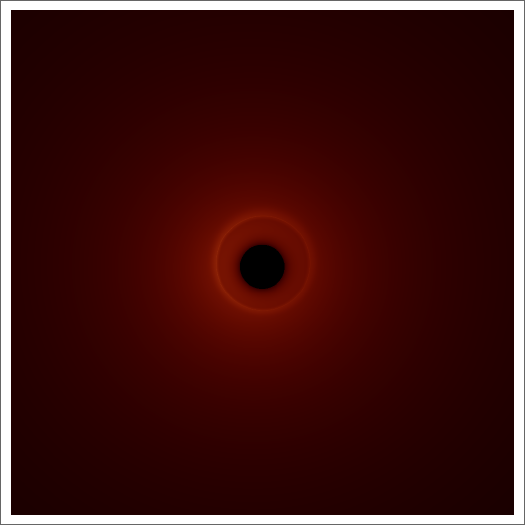}}
\subfigure[]{\includegraphics[width=.035\textwidth]{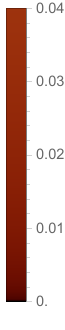}}
\subfigure[$a=0.998,\lambda=0.01$]{\includegraphics[width=.21\textwidth]{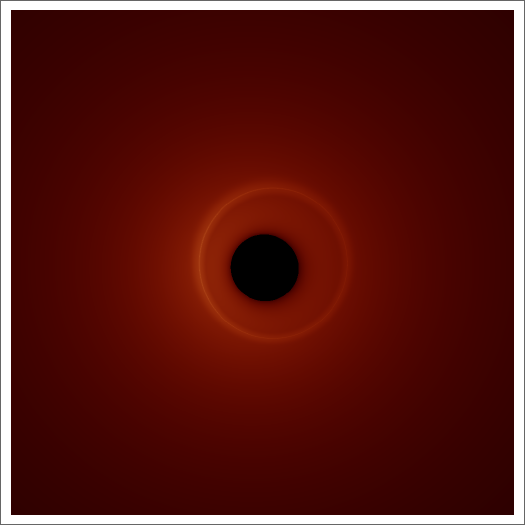}}
\subfigure[]{\includegraphics[width=.035\textwidth]{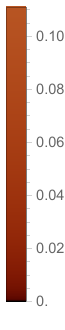}}
\subfigure[$a=0.998,\lambda=0.1$]{\includegraphics[width=.21\textwidth]{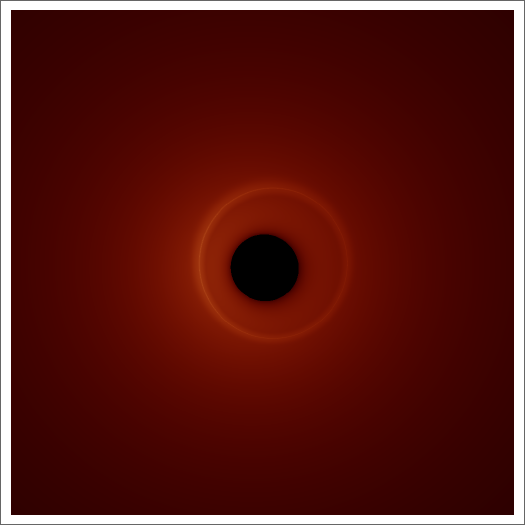}}
\subfigure[]{\includegraphics[width=.035\textwidth]{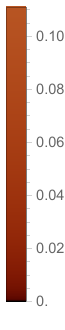}}}

\caption{\label{figbp} The images of the non-singular rotating black holes illuminated by the prograde flows.}
\end{figure}

From Fig.\ref{figbp}, when $\theta_{obs}=80^\circ$, the direct image and the lensed image of black hole are distinguishable; however, at $163^\circ$, these images become indistinguishable. By comparing with the contour of black hole's shadow, the inner shadow (for $a=0.1$) at $80^\circ$ shows a more noticeable deformation, appearing as a smooth, small semicircular black region. At $163^\circ$, the inner shadow remains a nearly circular black disk. For the two observed positions ($80^\circ$ and $163^\circ$), the image's brightness is symmetric vertically but distorted horizontally. When the observer is at $180^\circ$ or $0^\circ$, the brightness is symmetric in both horizontal and vertical directions. This is due to Doppler effects on the left side of the screen caused by the forward rotation of the prograde accretion disk. The increase of the rotation parameter directly results in a smaller inner shadow (at $163^\circ$) and deforms its shape (at $80^\circ$). On the other hand, the increase of the quantum parameter $\lambda$ causes the inner shadow of black hole to shrink in a manner that appears almost smooth and linear. More importantly, this parameter further reduces the observed intensity flux of the black hole. Due to the constraints on the parameter $\lambda$, the reduction in the shadow, inner shadow, and observed intensity is much smaller for high spin black hole than for the low spin case. In view of this, one can see that these results provide an effective tool to distinguish between nonsingular black holes in LQG from the Kerr black hole.

\begin{figure}[htp]
\makeatletter
\renewcommand{\@thesubfigure}{\hskip\subfiglabelskip}
\makeatother
\centering % \begin{center}/\end{center} takes some additional vertical space
\subfigure[(a) The redshifts of direct images of the accretion disk for $\theta_{obs}=80^\circ$.]{
\setcounter{subfigure}{0}
\subfigure[$a=0.1,\lambda=0.09$]{\includegraphics[width=.215\textwidth]{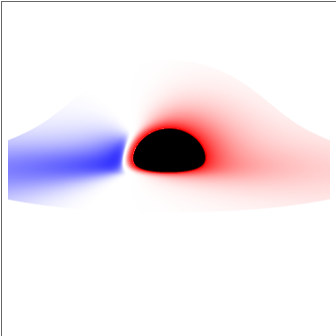}}
\subfigure[$a=0.1,\lambda=0.9$]{\includegraphics[width=.215\textwidth]{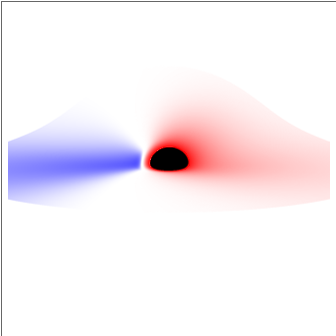}}
\subfigure[$a=0.998,\lambda=0.01$]{\includegraphics[width=.215\textwidth]{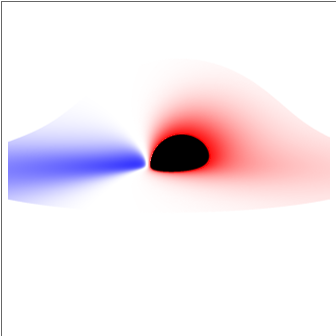}}
\subfigure[$a=0.998,\lambda=0.1$]{\includegraphics[width=.215\textwidth]{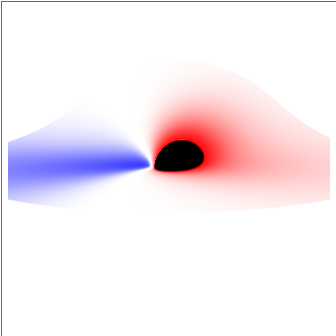}}
\subfigure[]{\includegraphics[width=.037\textwidth]{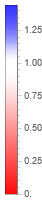}}}

\subfigure[(b) The redshifts of lensed images of accretion disk for $\theta_{obs}=80^\circ$.]{
\setcounter{subfigure}{0}
\subfigure[$a=0.1,\lambda=0.09$]{\includegraphics[width=.215\textwidth]{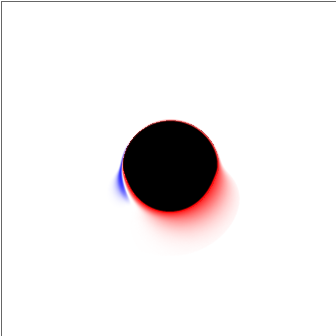}}
\subfigure[$a=0.1,\lambda=0.9$]{\includegraphics[width=.215\textwidth]{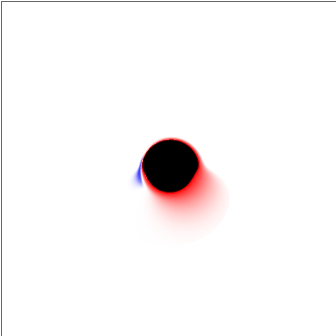}}
\subfigure[$a=0.998,\lambda=0.01$]{\includegraphics[width=.215\textwidth]{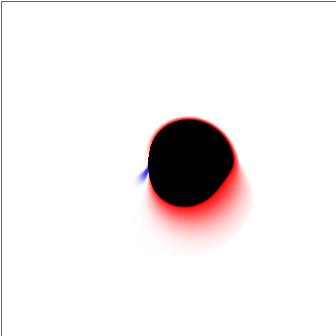}}
\subfigure[$a=0.998,\lambda=0.1$]{\includegraphics[width=.215\textwidth]{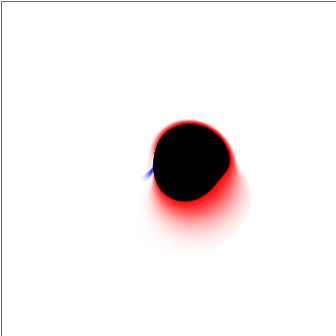}}
\subfigure[]{\includegraphics[width=.033\textwidth]{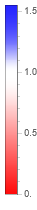}}}

\subfigure[(c) The redshifts of direct images of accretion disk $\theta_{obs}=163^\circ$.]{
\setcounter{subfigure}{0}
\subfigure[$a=0.1,\lambda=0.09$]{\includegraphics[width=.215\textwidth]{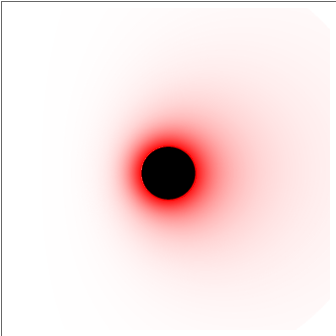}}
\subfigure[$a=0.1,\lambda=0.9$]{\includegraphics[width=.215\textwidth]{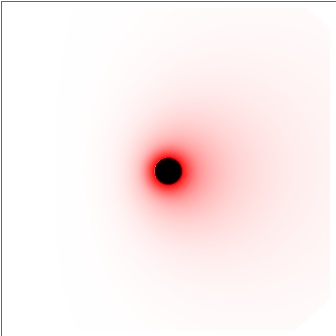}}
\subfigure[$a=0.998,\lambda=0.01$]{\includegraphics[width=.215\textwidth]{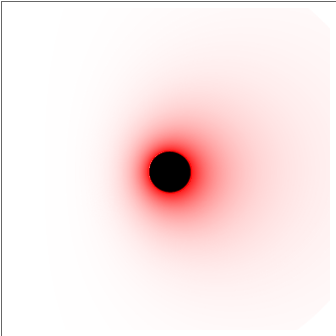}}
\subfigure[$a=0.998,\lambda=0.1$]{\includegraphics[width=.215\textwidth]{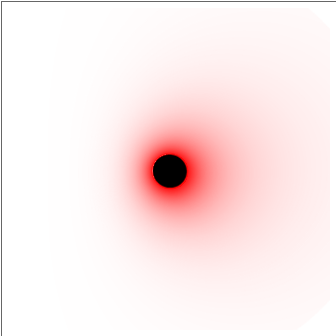}}
\subfigure[]{\includegraphics[width=.033\textwidth]{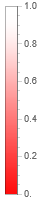}}}

\subfigure[(d) The redshifts of lensed images of accretion disk $\theta_{obs}=163^\circ$.]{
\setcounter{subfigure}{0}
\subfigure[$a=0.1,\lambda=0.09$]{\includegraphics[width=.215\textwidth]{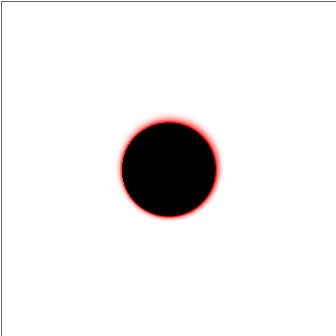}}
\subfigure[$a=0.1,\lambda=0.9$]{\includegraphics[width=.215\textwidth]{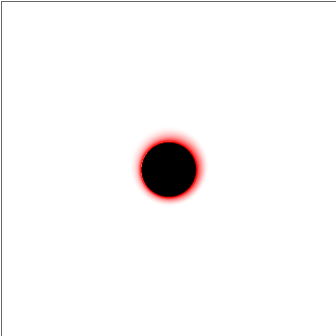}}
\subfigure[$a=0.998,\lambda=0.01$]{\includegraphics[width=.215\textwidth]{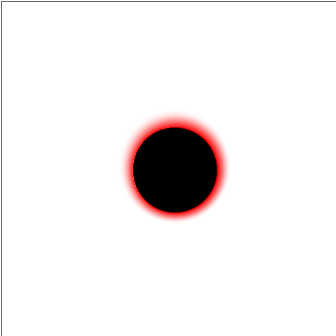}}
\subfigure[$a=0.998,\lambda=0.1$]{\includegraphics[width=.215\textwidth]{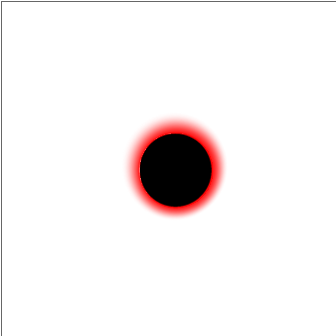}}
\subfigure[]{\includegraphics[width=.033\textwidth]{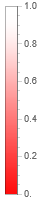}}}
\caption{\label{fighy1} The redshifts of direct and lensed images of the accretion disk model. }
\end{figure}

For $\theta_{obs}=80^\circ$, one can see from the above figures that when the rotation parameter increases, the transformation of the inner shadow results in noticeable changes in redshift. For example, at $80^\circ$, the increase of $a$ further deforms the semicircular inner shadow into an irregular shape. For $\theta_{obs}=163^\circ$, since the observer is relatively close to the North Pole, this effect is very weak.
More importantly, the quantum parameter $\lambda$ reduces the size of the black hole inner shadow, which is consistent with the previous discussion in the text regarding the black hole shadow.
Moreover, this parameter also has a substantial effect on the redshift and blueshift of black hole image. In particular, the parameter $\lambda$ decrease the range of the redshift and blueshift for both $a=0.1$ and $a=0.998$.
Meanwhile, one can see that the quantum parameter also reduces the shadow region clearly.
But for the lensed image in Fig.\ref{fighy1}(d), although $\lambda$ reduces the size of the shadow, it noticeably enhances the redshift of the lensed image at the edge of the black disk.
In view of this, it is true that the rotation parameter
$a$ and the quantum parameter $\lambda$ of the non-singular rotating black hole are very important for the redshift and blueshift images.

To carefully study the effect of $\lambda$ on the images of non-singular rotating black holes, we also plot the intensity distribution of the image, i.e., the intensity distribution along the X-axis and Y-axis.

\begin{figure}[htp]
\makeatletter
\renewcommand{\@thesubfigure}{\hskip\subfiglabelskip}
\makeatother
\centering % \begin{center}/\end{center} takes some additional vertical space
\subfigure[(a) The position of observer $\theta_{obs}=80^\circ.$]{
\setcounter{subfigure}{0}
\subfigure[$a=0.1$]{\includegraphics[width=.24\textwidth]{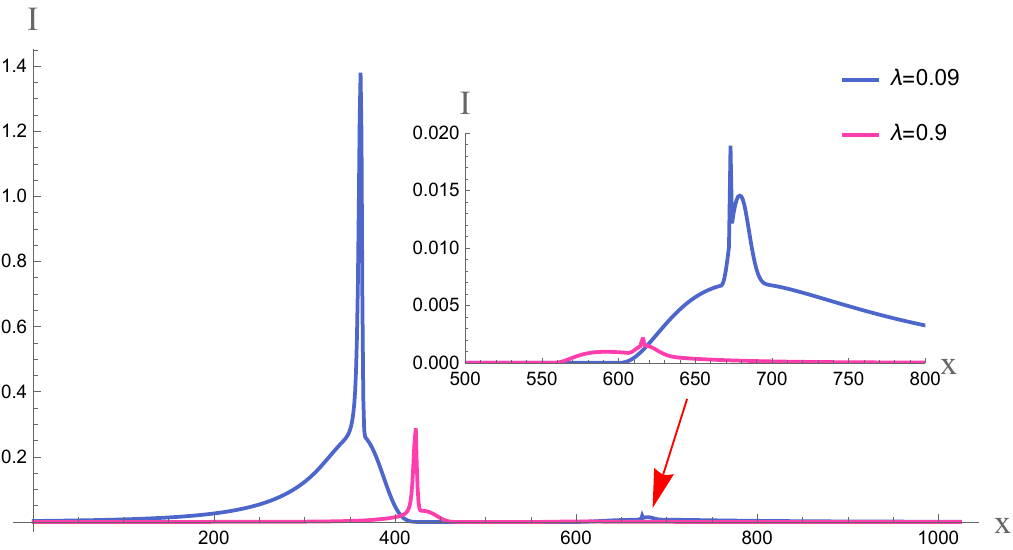}}
\subfigure[$a=0.1$]{\includegraphics[width=0.24\textwidth]{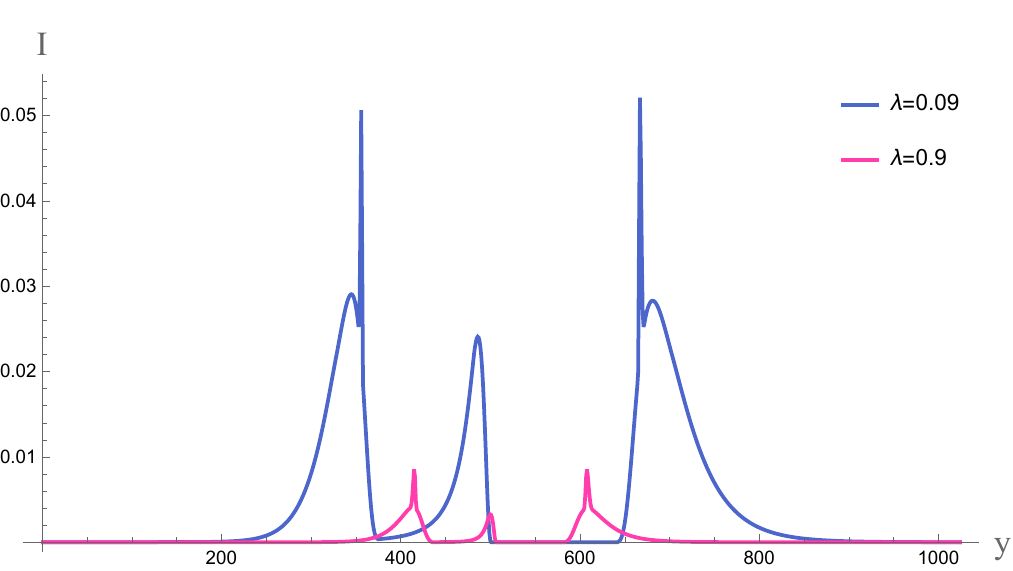}}
\subfigure[$a=0.998$]{\includegraphics[width=0.24\textwidth]{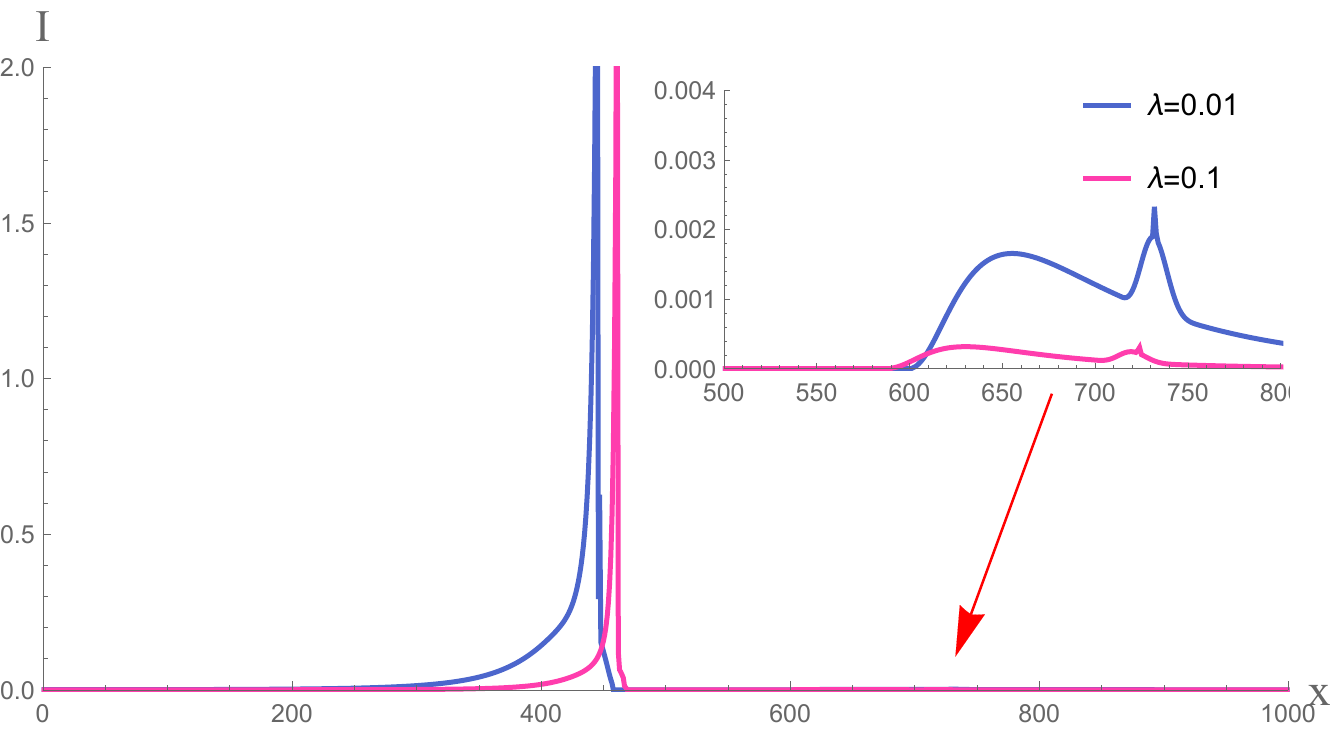}}
\subfigure[$a=0.998$]{\includegraphics[width=0.24\textwidth]{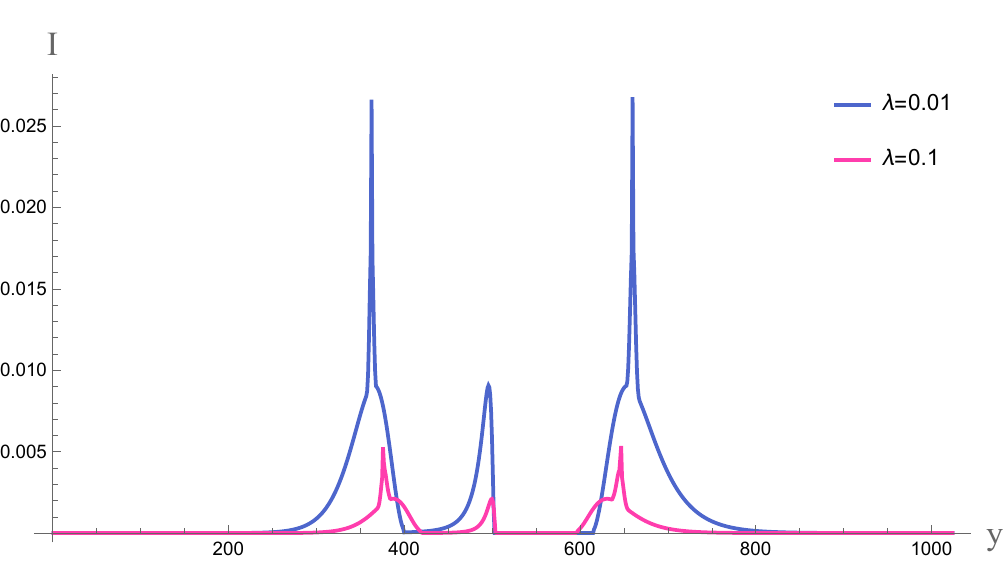}}}
\subfigure[(b) The position of observer $\theta_{obs}=163^\circ$]{
\setcounter{subfigure}{0}
\subfigure[$a=0.1$]{\includegraphics[width=0.24\textwidth]{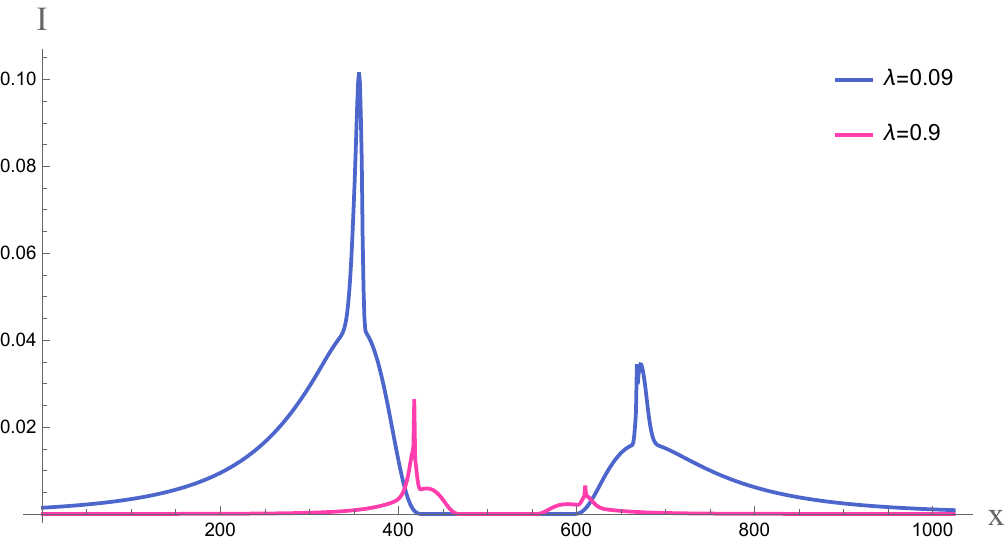}}
\subfigure[$a=0.1$]{\includegraphics[width=0.24\textwidth]{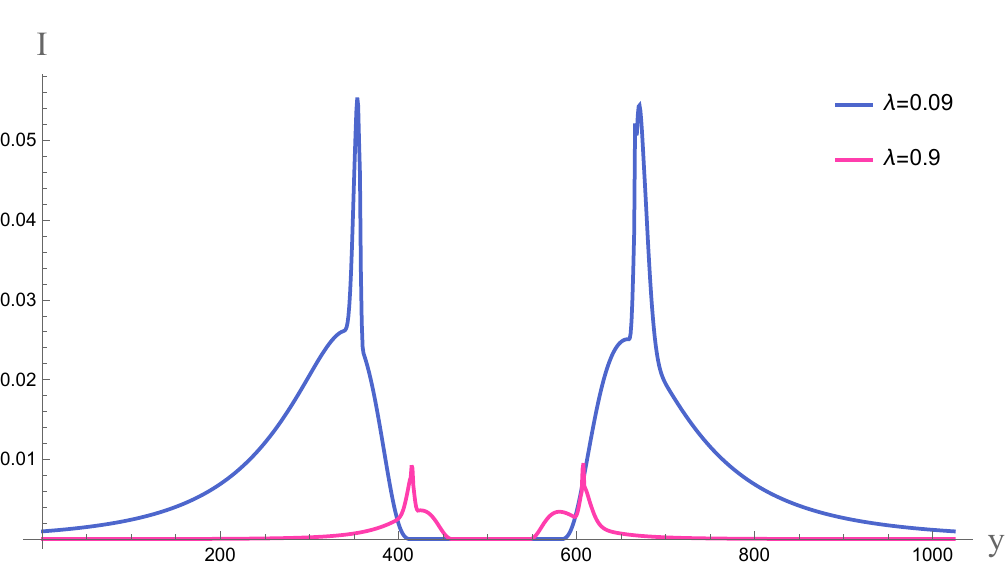}}
\subfigure[$a=0.998$]{\includegraphics[width=0.24\textwidth]{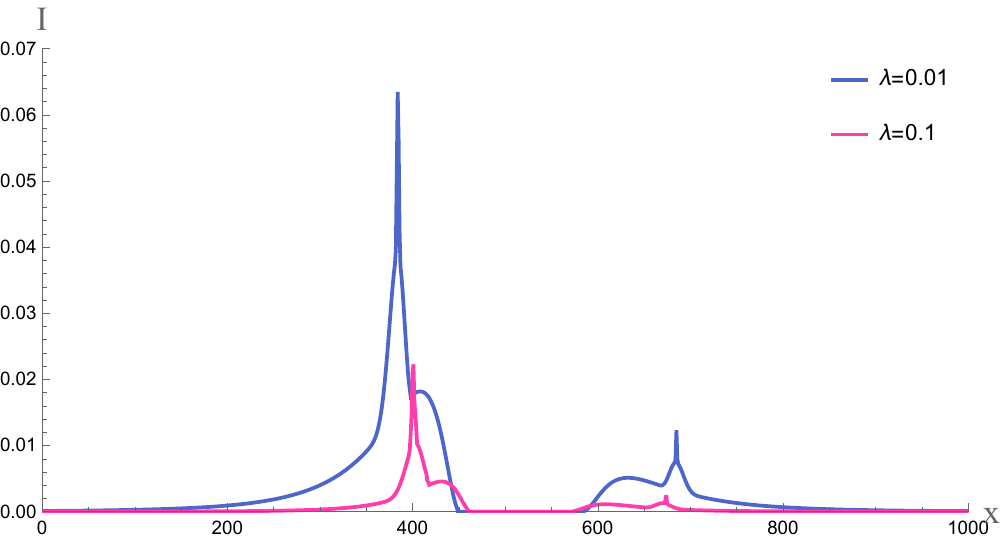}}
\subfigure[$a=0.998$]{\includegraphics[width=0.24\textwidth]{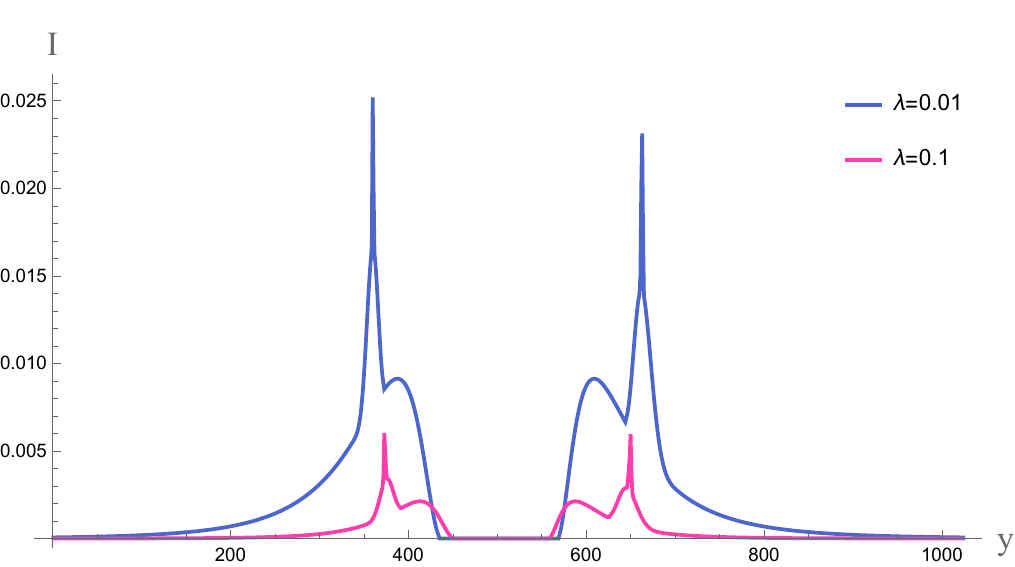}}}

\caption{\label{figbp} Intensity distribution along X-axis and Y-axis on the screen for the prograde flows.}
\end{figure}

At the position $80^\circ$, it shows in Fig.\ref{figbp} that there is a noticeable asymmetry in the intensity distribution along the X-axis on the screen, while the Y-axis remains relatively symmetric. The rotation parameter $a$ causes some variation in the peak intensity along the X-axis, with corresponding differences in the X-coordinates. More significantly, the quantum parameter has a pronounced impact on the intensity distribution. Whether along the X-axis or Y-axis, and whether $a=0.1$ or $a=0.998$, an increase in
$\lambda$ leads to a reduction in both the intensity and its peak value, along with a decrease in the spacing of the coordinates corresponding to the peaks. Moreover, this also holds true for the case at $163^\circ$.
Combining with the above analysis, we can conclude that, compared to Kerr black holes, the quantum parameter $\lambda$ of non-singular rotating black holes in LQG weakens the strength of gravitational field. This results in a reduction of the black hole shadow, a smaller inner shadow, and a reduction in the intensity of the observable thin disk image.

For the retrograde case, we present the accretion disk images, redshift images, and intensity distribution on the screen by using ($a=0.998,\lambda=0.01$) and ($a=0.998,\lambda=0.1$) as examples.

\begin{figure}[htp]
\makeatletter
\renewcommand{\@thesubfigure}{\hskip\subfiglabelskip}
\makeatother
\centering % \begin{center}/\end{center} takes some additional vertical space
\subfigure[(a) The images and intensity for $\theta_{obs}=80^\circ$ and $a=0.
998$.]{
\setcounter{subfigure}{0}
\subfigure[$\lambda=0.01$]{\includegraphics[width=.24\textwidth]{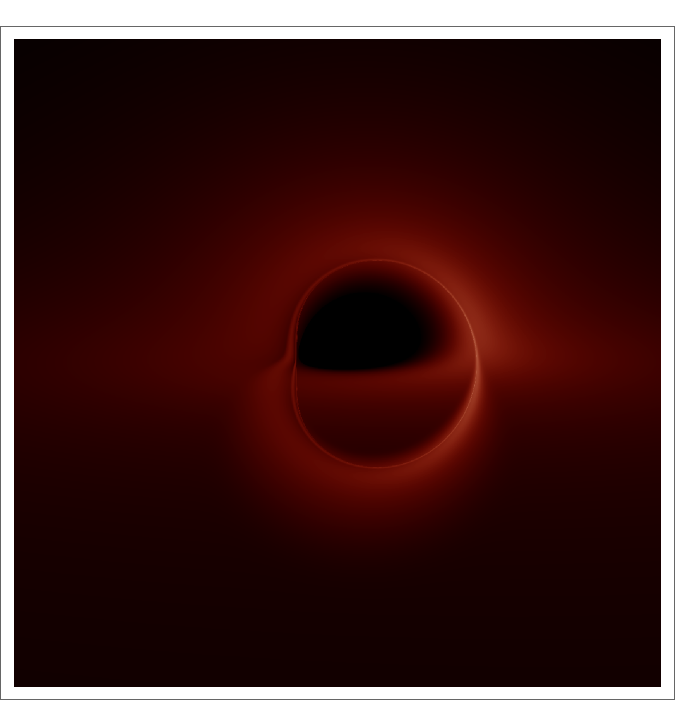}}
\subfigure[$\lambda=0.1$]{\includegraphics[width=0.24\textwidth]{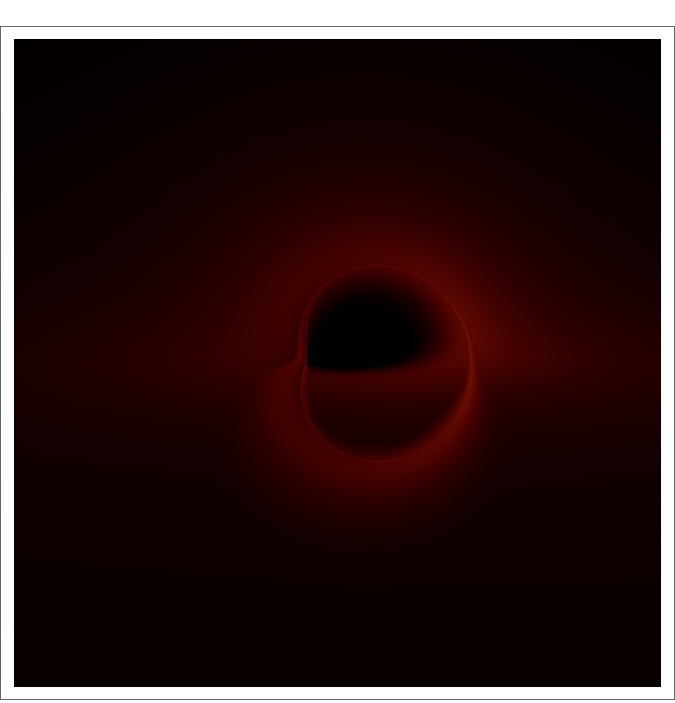}}
\subfigure[$\lambda=0.01$]{\includegraphics[width=0.24\textwidth]{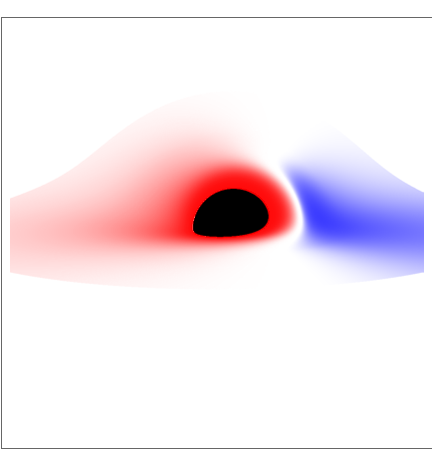}}
\subfigure[$\lambda=0.1$]{\includegraphics[width=0.24\textwidth]{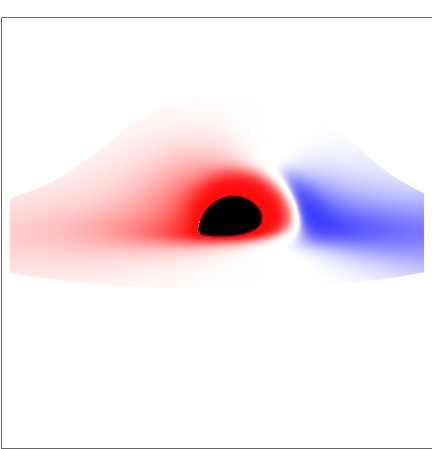}}}
\subfigure[(b) The position of observer $\theta_{obs}=80^\circ$]{
\setcounter{subfigure}{0}
\subfigure[$X-axis$]{\includegraphics[width=0.44\textwidth]{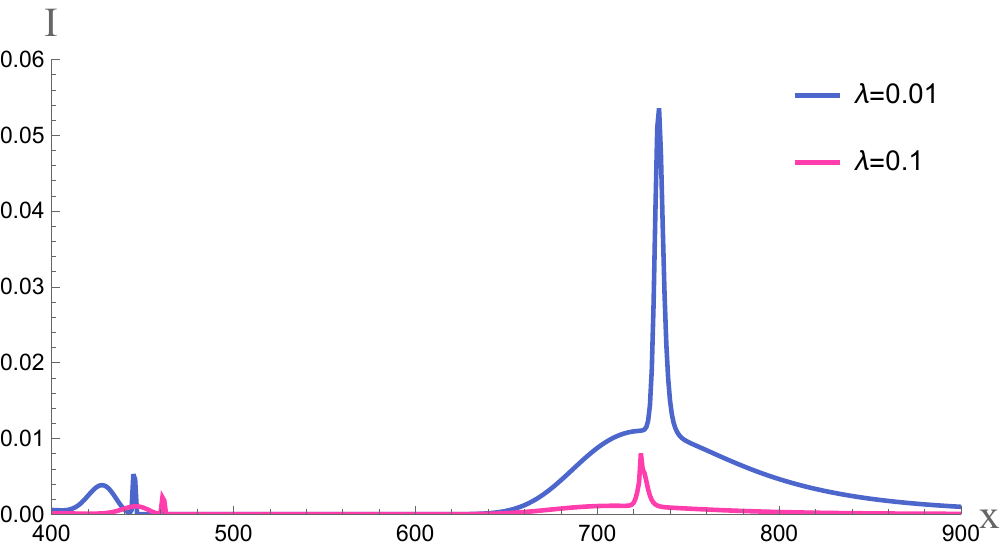}}
\hspace{1.8cm}
\subfigure[$Y-axis$]
{\includegraphics[width=0.44\textwidth]{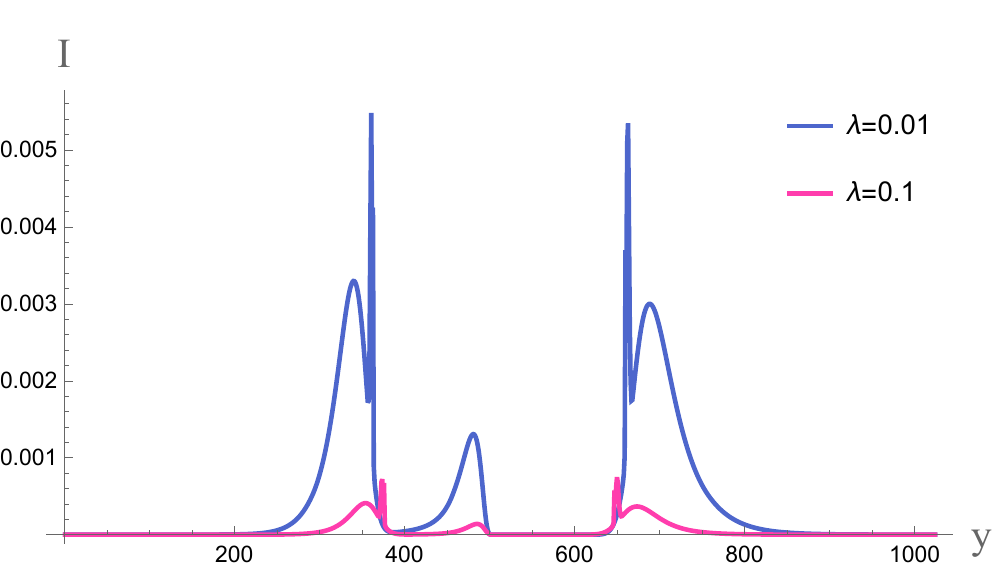}}}

\caption{\label{figbpnx} Intensity distribution along X-axis and Y-axis on the screen for the prograde flows.}
\end{figure}

Based on Fig.\ref{figbpnx}, we can observe that, under the same parameters, the inner shadow of the black hole remains unchanged. However, compared to the prograde scenario, the redshift and blueshift regions are reversed, with the redshift region around the shadow slightly expanding. Additionally, the maximum intensity of the light peak on the screen is now located on the right side of the shadow, rather than on the left as in the prograde case. Furthermore, the effects of the quantum parameter $\lambda$ on the shadow, redshift, and light intensity are consistent with those in the prograde scenario.

For the non-singular rotating spacetime in LQG, as shown in Fig.\ref{fig1}, the astronomical observations have ruled out the possibility of a rotating wormhole without an event horizon. Therefore, two possible scenarios remain for this rotating spacetime: first, a black hole with only one event horizon (i.e., the outer horizon), which implies that the transition surface encloses the inner horizon; second, a black hole with both an outer and an inner horizon, with the transition surface lying within the inner horizon. These two types of black holes exhibit different internal structures. We will continue to investigate whether these distinct internal structures produce any observable effects that can clearly distinguish between the two, or if there is any discontinuous behavior near the critical curve (the red curve in Fig.\ref{fig1}) as a result of these differences. Based on Fig.1, when $a=0.94$, $\lambda=0.0538$ lies precisely on the critical curve; $\lambda=0.0537$ corresponds to a non-singular rotating black hole with both an inner and outer horizon existed; and
$\lambda=0.0539$ represents a rotating spacetime with only an outer horizon. Taking the prograde disk as an example, we explore the observable appearance of these three scenarios at an inclination of $80^\circ$.

\begin{figure}[htp]
\makeatletter
\renewcommand{\@thesubfigure}{\hskip\subfiglabelskip}
\makeatother
\centering % \begin{center}/\end{center} takes some additional vertical space
\subfigure[(a) The images for $\theta_{obs}=80^\circ$ and $a=0.94.$]{
\setcounter{subfigure}{0}
\subfigure[$\lambda=0.357$]{\includegraphics[width=.3\textwidth]{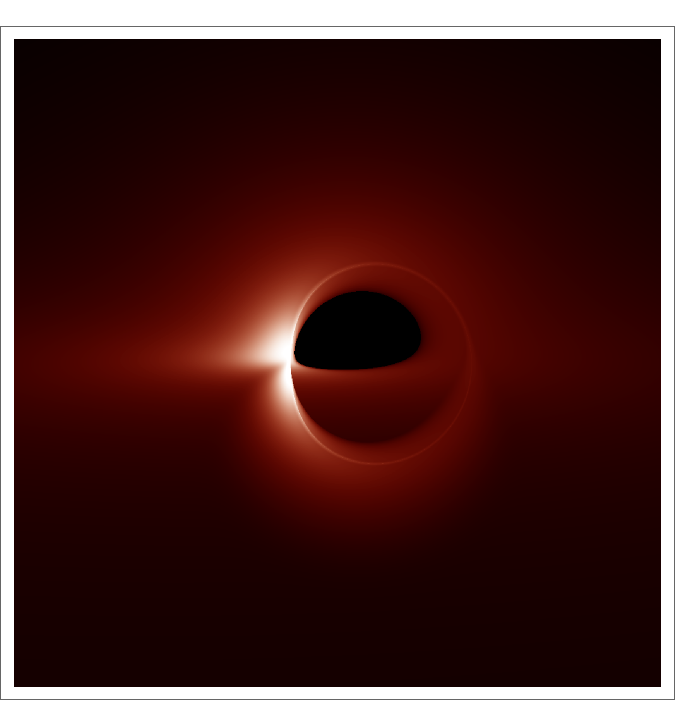}}
\subfigure[$\lambda=0.358$]{\includegraphics[width=0.3\textwidth]{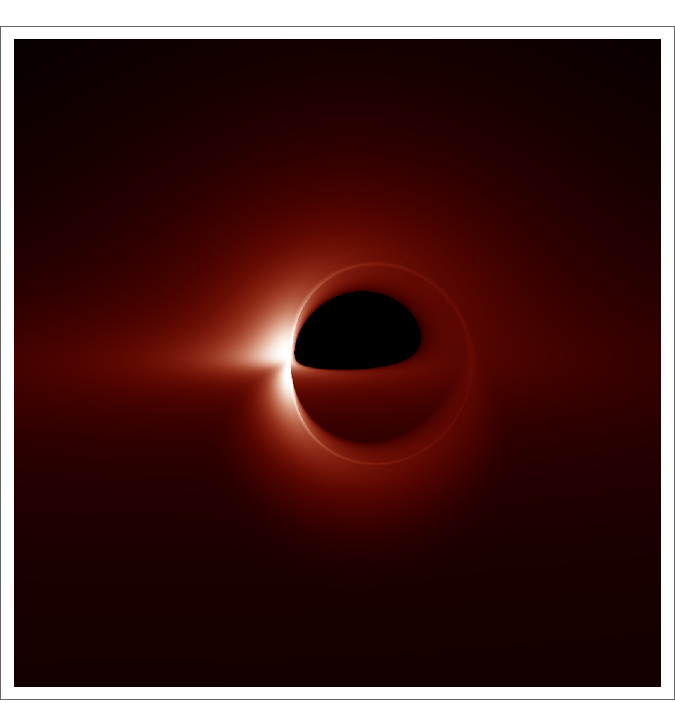}}
\subfigure[$\lambda=0.359$]{\includegraphics[width=0.3\textwidth]{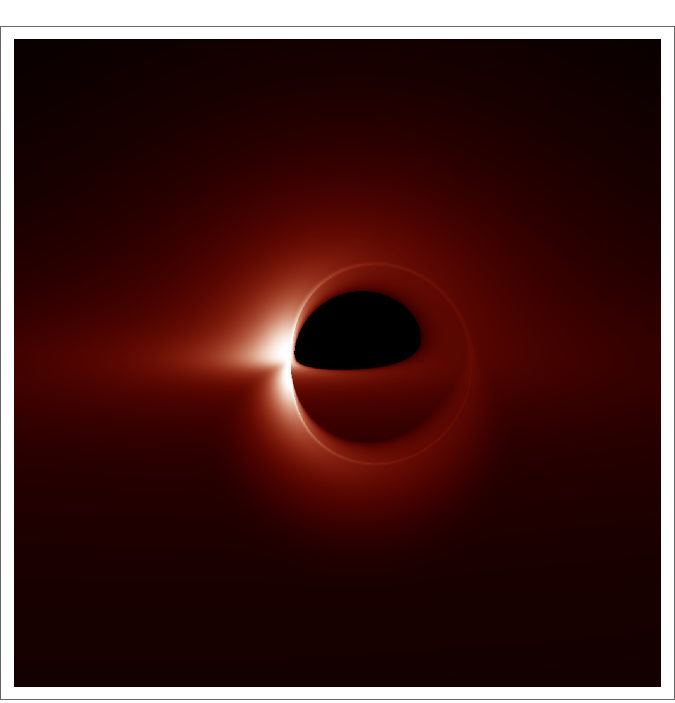}}}

\subfigure[(b) The redshift of derect images for $\theta_{obs}=80^\circ$ and $a=0.94$]{
\setcounter{subfigure}{0}
\subfigure[$\lambda=0.357$]{\includegraphics[width=0.3\textwidth]{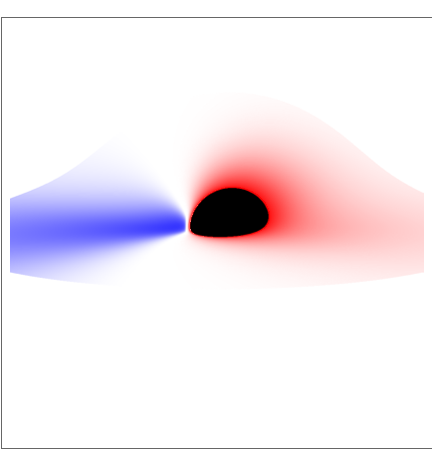}}
\subfigure[$\lambda=0.358$]
{\includegraphics[width=0.3\textwidth]{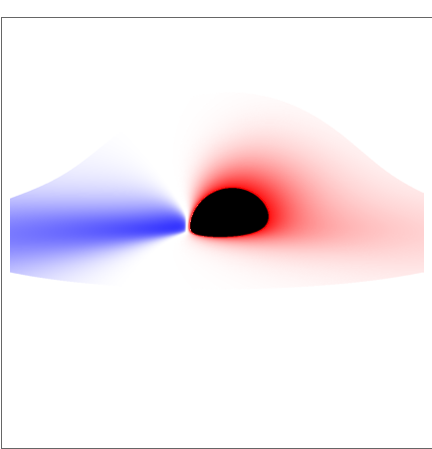}}
\subfigure[$\lambda=0.359$]
{\includegraphics[width=0.3\textwidth]{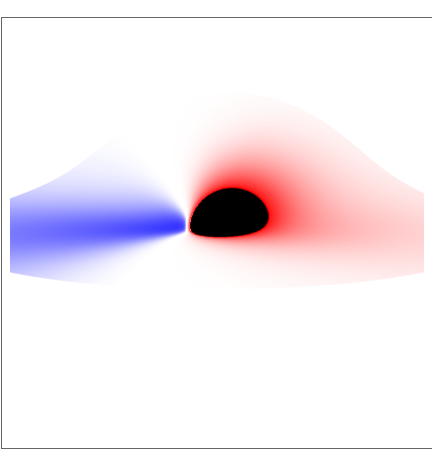}}}

\subfigure[(b) The intensity of images for  $\theta_{obs}=80^\circ$ and $a=0.94$.]{
\setcounter{subfigure}{0}
\subfigure[Y-axis]{\includegraphics[width=0.45\textwidth]{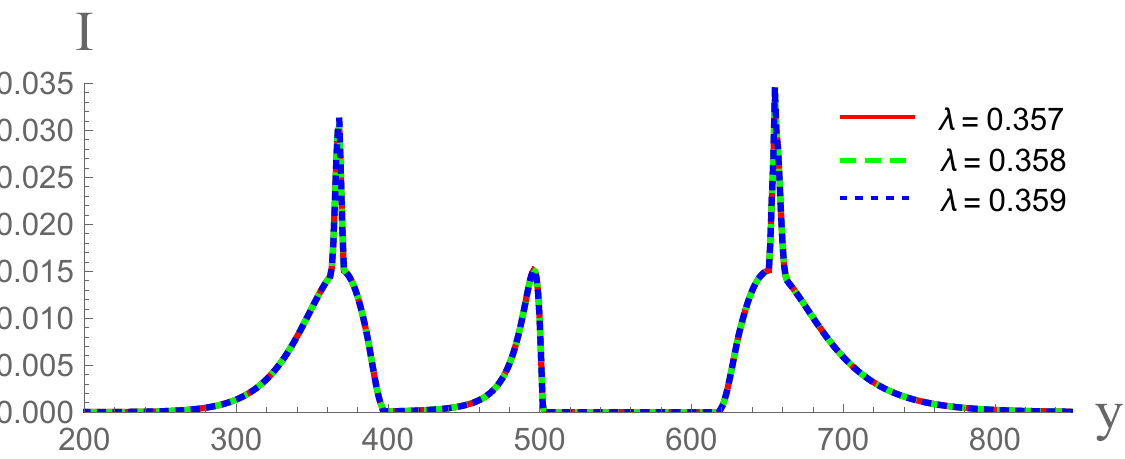}}
\hspace{0.2cm}
\subfigure[X-axis]
{\includegraphics[width=0.45\textwidth]{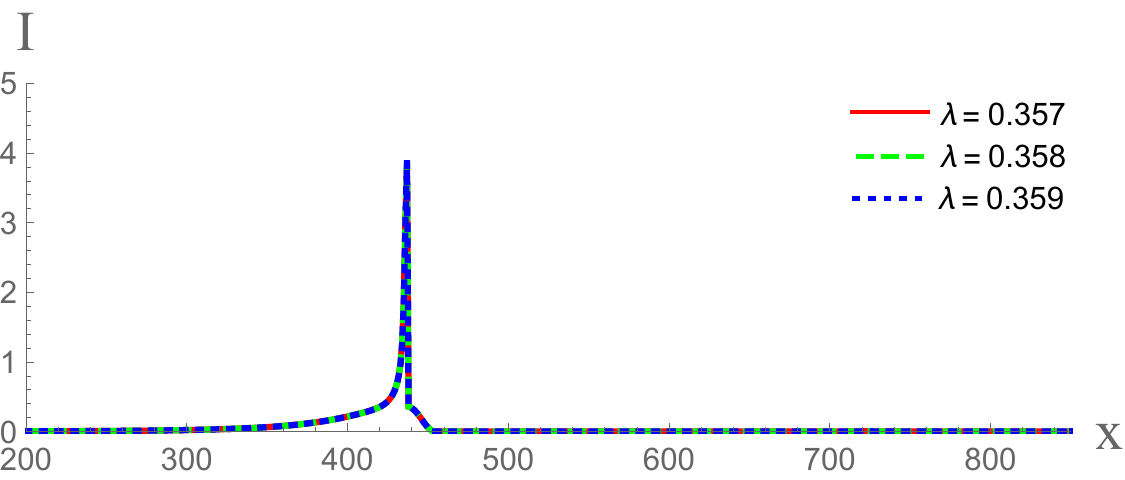}}}

\caption{\label{figbpdb} Intensity distribution along X-axis and Y-axis on the screen for the prograde flows.}
\end{figure}

From Fig.\ref{figbpdb}, for both the accretion disk image and its redshift profile, it is nearly impossible to discern whether the black hole has one horizon, two horizons, or lies on the critical curve. To carefully examine the differences in light intensity on the screen, we present the light intensity distributions and their peak values in the $X$ and $Y$ directions for all three cases. The results show that the intensity distributions in both the $X$ and $Y$ directions completely overlap, with almost no differences.
From this, we can conclude that the thin disk observational images of non-singular rotating black holes in LQG do not exhibit any noticeable discontinuous behavior near the critical curve. This implies that the thin disk images do not effectively reveal the internal structure of black holes, specifically the number of horizons. Moreover, the influence of the quantum parameter $\lambda$ on the image appears to manifest primarily through a reduction in the gravitational field's strength, and this effect seems to be linear. Obviously, whether this effect holds for thick accretion disks or jets remains unknown, and this requires further in-depth investigation.

\section{Conclusions and discussions}\label{sec7}
In this paper, based on the ray-tracing method, we investigate the shadows of non-singular rotating black holes in LQG, and further conduct numerical simulations of the inner shadows, celestial sphere images, and thin disk images of these black holes.
The non-singular rotating spacetimes encompasses three scenarios: (Region A) the rotating wormhole without the event horizon; (Region B) the black hole with only an outer event horizon; and (Region C) the rotating black hole with both an outer and inner event horizons. Since the astronomical observations have ruled out the possibility of a rotating wormhole, we focus on scenarios (Region B) and (Region C) for our study.
Firstly, we studied the shadow of non-singular rotating black holes using stereographic projection techniques in the ZAMO frame, and constrained the relevant parameters of these black holes using astronomical observation datas from M87* and SgrA*. Then, by considering a celestial sphere light source model, we continue to investigate the images of celestial sphere light sources and the behavior of light distortion near the black hole. Finally, by assuming the presence of an optically thin geometrical accretion disk in the equatorial plane of black hole, we also studied the characteristics of the inner shadow, thin disk images, and redshift profiles of black hole under both the prograde and retrograde disks.

The results show that the black hole shadow is closely related to the observer's position ($r_{obs}$, $\theta_{obs}$) and black hole's parameters ($a$, $\lambda$). The farther the observer is, the smaller the shadow; the smaller the observer's angle ($\theta_{obs} \in [0,\pi/2]$), the closer the shadow is to a circular shaped. As the spin parameter increases, the shadow observed from the equatorial plane becomes more ``D"-shaped, while the increase of the quantum parameter $\lambda$ causes the shadow to shrink.
So, the parameter $\lambda$ decreases the size of the shadow, but increases the deviation from circularity.
Moreover, when ($a=0.5$), the range of $\lambda$ is $\in$ $[0, 0.62]$. Based on the datas from M87* and SgrA*, we find that the entire range of $\lambda$ falls within the permissible range for the angular diameter of SgrA*'s shadow.
However, to ensure the shadow's angular diameter of M87* falling within its 1$\sigma$ confidence interval, it requires $\lambda$ $\in$ $[0, 0.37]$, and being within the 2$\sigma$ confidence interval, it requires $\lambda$ $\in$ $[0, 0.52]$. This indicates that the observations of M87* impose stronger constraints on the observed properties of the no-singulgar rotating black holes by comparing with SgrA*.
When an ideal celestial light source illuminated the black hole, it shows that there is the significant distortion of light around black hole, and the shadow gradually transforms into a ``D" shape as $a$ increases. Meanwhile, as the parameter $\lambda$ increases, the shadow gradually shrinks. On the surface, it seems that $\lambda$ only affects the size of the shadow. However, after examining it closely, it becomes apparent that different values of $\lambda$ lead to the distinct behavior of the light near the left side of black hole shadow, as seen in the changes in the blue and green regions. This indicates that the quantum parameter $\lambda$  indeed influences the distortion of light around black hole shadow, although this effect is relatively small and only becomes noticeable near the event horizon.

When the thin accretion disk is located at the equatorial plane with the prograde angular velocity, for an observer with $\theta_{obs} = 80\pi/180$, the results show that the spin parameter $a$ plays a dominant role in shaping the inner shadow of black hole, and its influence on the size of the inner shadow is relatively weak. On the other hand, the quantum parameter $\lambda$ has little effect on the shape of black hole shadow but plays a dominant role in determining the size of the inner shadow. Moreover, for the image of non-singular rotating black holes, the direct image and the lensed image are distinguishable.
When the observer is located at $163^\circ$, the inner shadow of black hole appears almost as a black disk, and at this point, the influence of $a$ and $\lambda$ on the black hole's shadow is reflected only in its size. Comparatively, the effect of
$\lambda$ is more pronounced. At the same time, the direct image and the lensed image are nearly indistinguishable at this inclination.
For the observed intensity, whether at $80^\circ$ or $163^\circ$, the intensity along the Y-axis is nearly symmetrical, while along the X-axis, it is stronger on the left and weaker on the right. Both $a$ and $\lambda$ cause the two peaks of the intensity to move closer together, though the influence of $\lambda$ is more pronounced. Furthermore, $\lambda$ has a more noticeable effect on the observed intensity, primarily reducing the intensity distribution area and its maximum value.
Additionally, it further shows that both
$a$ and $\lambda$ mainly decrease the size of the shadow of redshift images, and reduce the area of the redshift images.
Furthermore, based on the analysis of retrograde accretion disk images and redshift images of non-singular rotating black holes, it is found that the influence of $a$ and $\lambda$ on the inner shadow, observed intensity, and redshift images remains consistent with that in the prograde case.
Finally, we attempted to examine whether the appearance of thin disk accretion for non-singular rotating black holes can reflect their different internal structures. When fixing $a=0.94$, $\lambda=0.0539$ corresponds to case (Region B), $\lambda=0.0537$ corresponds to case (Region C), and $\lambda=0.0538$ is at the critical value. It turns out that the black hole's accretion disk images, redshift images, and intensity distribution curves on the screen in all three cases reveal no discernible differences. This indicates that the thin disk accretion images of black hole cannot effectively reflect the internal structure of black hole.

Based on the above facts, we can conclude that a key observational feature of these non-singular rotating black holes is that the larger the black hole's spin parameter, the smaller the upper limit of the quantum parameter $\lambda'$s effect.
This effect manifests as, the quantum parameter $\lambda$ decreases the gravitational field's strength, shrinks of the shadow size, increases the circular deviation of shadow, reduce the inner shadow, weakens the observed intensity in thin disk images, attenuates the shadow in redshift images, and shrinks of the shadow area. And, this effect is more obvious than that of the rotating parameters $a$.
Although the thin disk accretion images do not reveal the internal structure of black hole, these characteristics of the $\lambda$'s effect differ significantly from those of Kerr black holes. This could provide a possible way to constraining black hole parameters, identifying QG effects, and distinguishing LQG black holes. Of course, it is also interesting to check this effect in thick disk accretion images, polarized images, or hotspot images in future studies.

\vspace{10pt}

\noindent {\bf Acknowledgments}

\noindent
This work is supported by the National Natural Science Foundation of China (Grant No.11903025), and by the Sichuan Science and Technology Program (2023ZYD0023), and by the Sichuan Youth Science and Technology Innovation Research Team (21CXTD0038).\\

\end{document}